\documentclass[final,5p,times,twocolumn]{elsarticle}

\usepackage{amssymb}
\usepackage{amsmath}
\usepackage{booktabs}
\usepackage{tabularx}
\usepackage{url}

\makeatletter
\long\def\@makecaption#1#2{%
  \vskip\abovecaptionskip\footnotesize
  \sbox\@tempboxa{#1. #2}%
  \ifdim \wd\@tempboxa >\hsize
    #1. #2\par
  \else
    \global\@minipagefalse
    \hb@xt@\hsize{\hfil\box\@tempboxa\hfil}%
  \fi
  \vskip\belowcaptionskip}
\makeatother

\journal{Journal of Parallel and Distributed Computing}

\begin{document}

\begin{frontmatter}



\title{A HIP-Compatible Accelerator Backend for Fourier-Bessel Particle-in-Cell Simulations on CPU/DCU Heterogeneous Clusters} %% Article title


\author[label1,label2]{Jingliang Fan}

\author[label3]{Ruiqing He}

\author[label4]{Yang Wan}

\author[label1]{Jiandong Shang}

\author[label1]{Hengliang Guo}

\author[label1,label5]{Qiang Chen\corref{cor1}}
\ead{qiangchen@zzu.edu.cn}

\cortext[cor1]{Corresponding author}

%% Author affiliations
\affiliation[label1]{
    organization={National Supercomputing Center in Zhengzhou, Zhengzhou University},
    city={Zhengzhou},
    postcode={450001},
    country={China}
}

\affiliation[label2]{
    organization={School of Computer and Artificial Intelligence, Zhengzhou University},
    city={Zhengzhou},
    postcode={450001},
    country={China}
}

\affiliation[label3]{
    organization={School of Communication and Artificial Intelligence, School of Integrated Circuits, Nanjing Institute of Technology},
    city={Nanjing},
    postcode={211167},
    country={China}
}

\affiliation[label4]{
    organization={School of Physics and Laboratory of Zhongyuan Light, Zhengzhou University},
    city={Zhengzhou},
    postcode={450001},
    country={China}
}

\affiliation[label5]{
    organization={Laboratory for Advanced Computing and Intelligence Engineering},
    city={Wuxi},
    postcode={214000},
    country={China}
}

%% Abstract
\begin{abstract}
FBPIC (Fourier-Bessel particle-in-cell) is a high-performance simulation code for relativistic plasma and accelerator physics. Its original accelerator backend relies on Numba CUDA, which limits its direct deployment on accelerators using the HIP (Heterogeneous-Compute Interface for Portability) programming environment, such as DCU (Deep Computing Unit) accelerators. In this work, we develop an accelerator backend compatible with HIP that enables FBPIC to run efficiently on DCU platforms while preserving its Python user interface and high level simulation workflow.  For the evaluated LWFA (laser-wakefield acceleration) workloads, the proposed backend achieves 1.32-1.54× speedups over the original FBPIC implementation on an NVIDIA V100 GPU and enables efficient execution on the DCU platform. We also summarize the key lessons learned from porting FBPIC to the DCU platform. Multi-DCU experiments achieve a 1.88× strong-scaling speedup on four accelerators and a 2.72× increase in aggregate throughput at approximately 68\% weak-scaling efficiency, with communication analysis identifying inter-node communication and synchronization as the main scalability limitations. Beyond FBPIC, the proposed approach provides a practical reference for porting and optimizing other scientific computing applications developed with Python on heterogeneous accelerator platforms.
\end{abstract}

%% Keywords
\begin{keyword}
    Fourier-Bessel particle-in-cell \sep HIP \sep CuPy \sep DCU \sep Performance portability
\end{keyword}

\end{frontmatter}



\section{Introduction}\label{sec1}
The PIC (Particle-In-Cell) method has been widely used for plasma physics, accelerator physics, and laser-plasma interaction simulations\cite{birdsall1991plasma, fonseca2002osiris, decyk2014architectures}. These problems involve electromagnetic structures and charged particles evolving over multiple spatial and temporal scales, and their predictive accuracy is often limited by both numerical dispersion and computational cost. As plasma-accelerator studies move toward larger domains, longer propagation distances, and more extensive parameter scans, accelerator computing has become increasingly important for high-fidelity PIC simulations\cite{decyk2014architectures, vay2018warpx}. The DCU (Deep Computing Unit) is a GPU-like accelerator for high-performance heterogeneous computing and supports the HIP (Heterogeneous-computing Interface for Portability) programming model\cite{liu2024dcu}. Its CUDA-like programming interface provides a practical environment for porting GPU-accelerated scientific applications\cite{amd2026hip,tsai2021ginkgo}. With the emergence of such accelerator platforms, improving the performance portability of high-fidelity PIC codes becomes increasingly important. FBPIC (Fourier-Bessel Particle-In-Cell) is a Particle-In-Cell code designed for relativistic plasma simulations, with particular relevance to laser-wakefield and plasma-wakefield acceleration problems\cite{lehe2016fbpic}. By representing the electromagnetic fields in a quasi-cylindrical spectral basis, FBPIC can retain essential three-dimensional physical effects for nearly axisymmetric systems while requiring far fewer degrees of freedom than a full three-dimensional Cartesian PIC simulation\cite{lehe2016fbpic}. The Fourier-Bessel spectral formulation also avoids the numerical dispersion associated with conventional finite-difference time-domain PIC solvers, which is particularly important for simulations involving ultra-relativistic beams and laser pulses\cite{lehe2016fbpic,godfrey2014psatd}. 

The original FBPIC implementation is written in Python and relies on Numba just-in-time compilation for its CPU and CUDA GPU execution paths. Its GPU support requires both CuPy and Numba-related GPU dependencies\cite{fbpic2026install,lam2015numba,okuta2017cupy}. This design provides a productive high-level programming model and good performance on NVIDIA GPUs, but it also tightly couples the performance-critical kernels to the CUDA-oriented Numba backend. This becomes a major limitation when targeting DCU/HIP platforms, because Numba’s ROCm target has been officially unmaintained since version 0.54.0\cite{numba2021rocm}. As a result, the original Numba-based GPU path cannot be directly reused as a robust and efficient backend for DCU systems.

To overcome this portability limitation, we develop a HIP-compatible accelerator backend for FBPIC that removes the dependence on Numba from its performance-critical GPU execution path. Instead of translating Python-defined kernels through the Numba JIT infrastructure, the dominant PIC kernels are reimplemented as explicit C/C++ GPU kernels and integrated into the existing Python framework through CuPy RawKernel\cite{cupy2026rawkernel}. CuPy provides CUDA- and ROCm-backed execution environments, allowing the same Python-side dispatch mechanism to be retained across NVIDIA GPUs and HIP-compatible accelerators\cite{okuta2017cupy,cupy2026rocm}. In addition, particle and field data remain resident in CuPy device arrays throughout the time-stepping loop, so the backend can be replaced without introducing additional host-device data movement or modifying the high-level simulation workflow. The existing MPI-based longitudinal domain decomposition is preserved, so the proposed backend changes only the intra-device execution layer rather than the physical model or distributed-memory decomposition of FBPIC. 

The main contributions of this work are summarized as follows: 

\noindent\textbf{A portable accelerator backend for FBPIC.}
We develop a HIP-compatible accelerator backend for FBPIC by replacing its performance-critical Numba CUDA kernels with explicitly implemented C/C++ GPU kernels. These kernels are integrated into the existing Python framework through CuPy RawKernel, preserving the original user interface, simulation workflow, device-resident data model, and MPI-based domain decomposition.

\noindent\textbf{Kernel-level optimization for DCU execution.}
We optimize the rewritten kernels for DCU execution through compiler- and occupancy-oriented techniques, including pointer-based memory access, selective loop unrolling, register-pressure control, and kernel-specific thread-block tuning, and investigate the performance trade-offs associated with particle locality and binning.

\noindent\textbf{Cross-platform numerical and performance evaluation.}
We verify the numerical consistency of the backend using linear-wakefield and  LWFA (laser-wakefield acceleration) benchmarks and demonstrate performance portability across CUDA and HIP platforms. On an NVIDIA V100 GPU, the rewritten implementation achieves 1.32-1.54× speedups over the original FBPIC implementation for the evaluated workloads.

\noindent\textbf{Multi-DCU scalability and communication analysis.}
We characterize the multi-DCU scalability of FBPIC through strong- and weak-scaling experiments and mpiP communication profiling. The implementation achieves a 1.88× strong-scaling speedup on four DCUs and a 2.72× increase in aggregate throughput at approximately 68\% weak-scaling efficiency, while the profiling results identify inter-node communication and synchronization as the primary limitations at larger process counts.

\section{Current state of the art}
\label{sec2}
%% Use \subsection commands to start a subsection.
Considerable effort has been devoted to improving the accuracy and scalability of PIC simulations. Conventional Cartesian finite-difference PIC solvers remain widely used because of their algorithmic locality and compatibility with domain decomposition. However, for relativistic beams and high-speed laser propagation, finite-difference field solvers may introduce numerical dispersion and numerical Cherenkov-type artifacts\cite{godfrey1974cherenkov,vay2011mitigation}. Spectral and pseudo-spectral solvers have therefore become important alternatives. FBPIC follows this direction by adopting a Fourier-Bessel quasi-cylindrical spectral formulation, which is particularly effective for close-to-axisymmetric laser wakefield and plasma wakefield acceleration problems. Its spectral cylindrical representation provides high accuracy at substantially lower cost than full three-dimensional Cartesian PIC for such geometries\cite{lehe2016fbpic}.

Modern high-performance PIC codes increasingly target many-core and GPU-based supercomputers. WarpX is an exascale-oriented electromagnetic PIC code built on the AMReX
framework. Its performance-portable implementation targets NVIDIA, AMD, and Intel GPUs and has demonstrated scalability on leadership-class systems\cite{myers2021warpx,fedeli2022warpx}. PIConGPU represents another major GPU-oriented development path. It was originally developed as a CUDA-based PIC implementation for GPU clusters\cite{burau2010picongpu}. It was subsequently moved toward performance portability using the cupla interface over Alpaka, enabling single-source C++ kernels to target different processor architectures\cite{zenker2016picongpu}. Smilei is a collaborative, open-source C++ PIC code co-developed by plasma physicists and HPC specialists, with an emphasis on modularity, multiphysics capabilities, and parallel performance\cite{derouillat2018smilei}. HiPACE++ further demonstrates the effectiveness of performance-portable GPU implementations for quasi-static plasma-accelerator modeling, reporting near-optimal strong scaling from 1 to 512 GPUs\cite{diederichs2022hipace}.
 
These developments illustrate a broader shift in high-performance PIC software toward accelerator-aware and performance-portable implementations\cite{myers2021warpx,zenker2016picongpu,diederichs2022hipace}. Representative implementations are predominantly developed in C++ and rely on portability layers such as AMReX or Alpaka to map computational kernels onto
different processor architectures\cite{zhang2021amrex,zenker2016picongpu}. FBPIC differs from these codes in that it combines a high-level Python workflow with a spectral quasi-cylindrical algorithm. The original implementation relies on Numba JIT compilation for performance and can run on multicore CPUs or GPUs, with GPU execution being much faster for large simulations\cite{fbpic2026docs}. This Python-JIT design provides productivity and flexibility, but it also limits portability. In particular, the original GPU backend is closely tied to Numba CUDA. This becomes problematic for AMD/DCU-class accelerators because Numba’s ROCm target was moved to an unmaintained status in version 0.54.0 and relocated outside the main Numba repository\cite{numba2021rocm}.

A recent experimental FBPIC pull request explored the use of the \texttt{numba.hip} interface with \texttt{numba.hip.pose\_as\_cuda()} in order to minimize changes to the existing CUDA-oriented implementation\cite{sinn2025fbpicamd}. The reported tests on one GPU die of an MI250X identified several interoperability and functionality limitations and showed lower-than-expected performance\cite{sinn2025fbpicamd}. These observations suggest that a minimal-change Numba-HIP compatibility layer
alone may be insufficient to provide a robust and high-performance FBPIC backend for production use.

CuPy provides another possible route for Python-based GPU computing. Its RawKernel interface allows user-defined kernels written in raw CUDA source to be compiled and launched from Python, while caching the compiled binary for reuse\cite{cupy2026rawkernel}. CuPy also provides experimental ROCm support. It provides a practical integration layer through which carefully written C/C++ GPU kernels can be connected to a Python scientific application.
The present work focuses on this unresolved gap. While preserving the usability of FBPIC, this work implements a version that can run efficiently across multiple GPU platforms by leveraging the RawKernel interface provided by CuPy. This study complements existing research on PIC code portability and provides a practical reference for the efficient and robust migration of other Python-based scientific computing applications to emerging GPU-like accelerator platforms such as DCUs.

\section{Parallel implementation of FBPIC on DCU accelerators}
\label{sec3}
\subsection{Computational structure and porting requirements of FBPIC}
\label{subsec:computational-structure}
FBPIC is organized as a Python-controlled, object-oriented simulation framework. The top-level Simulation object owns the field, particle, diagnostic, and boundary-communication components and invokes the PIC cycle through its step method. The high-level Python layer is primarily responsible for physical configuration, data organization, and execution control, whereas the computationally intensive operations are implemented as device kernels or GPU array operations. In the original implementation, GPU execution requires both CuPy and Numba CUDA\cite{fbpic2026install}. FBPIC exposes two levels of parallelism. At the inter-device level, the computational domain is decomposed along the longitudinal direction, and neighboring MPI ranks exchange field guard cells and particles that cross subdomain boundaries. At the intra-device level, particles and grid elements within each subdomain are processed concurrently by GPU threads\cite{fbpic2026parallel,jalas2017psatd}. The present work preserves the existing longitudinal domain decomposition and MPI-level simulation semantics, while replacing the intra-device Numba implementation and the associated buffer-processing kernels. This separation limits platform-specific modifications to the accelerator backend and avoids changes to the physical decomposition of the original code. The five major computational stages of a typical FBPIC time step are described below, and their key characteristics are summarized in Table~\ref{tab:chara}.

Field gathering. The electric and magnetic fields stored on the interpolation grids are evaluated at the particle positions. Because FBPIC represents the fields as a truncated set of azimuthal modes, the local field experienced by each particle is reconstructed by accumulating contributions from all retained modes. Gathering is dominated by irregular read accesses and repeated interpolation operations.

Particle push. After gathering, the particle momenta and positions are advanced according to the Lorentz force. FBPIC uses a time-centered PIC sequence in which the momentum and position updates are staggered relative to the field quantities. The update of each macroparticle is independent, which provides abundant data parallelism.

Charge and current deposition. Particle charge and current are projected back to the interpolation grids. Similar to gathering, each particle contributes to several radial and longitudinal grid points for every azimuthal mode. Unlike gathering, deposition performs concurrent updates to shared grid locations. Multiple particles may therefore write to the same element, requiring atomic accumulation or an equivalent conflict-resolution strategy.

Spectral source correction and field advancement. The deposited sources are transformed to spectral space, corrected to satisfy the selected current-continuity treatment, and used to advance Maxwell's equations with the pseudo-spectral analytical time-domain method. The spectral representation combines Fourier transforms along $z$ with Hankel transforms along $r$. This spectral field-solver formulation avoids the numerical dispersion commonly associated with conventional finite-difference field solvers.

Boundary and particle exchange. For multi-device simulations, guard-cell values are exchanged between adjacent longitudinal subdomains, deposited source quantities are accumulated across overlapping regions, and particles leaving a local subdomain are transferred to the neighboring rank. The communication volume is determined by the guard-region width and the number of migrating particles rather than the total local particle population. Local packing, unpacking, buffer initialization, and particle rearrangement remain accelerator-intensive operations and must therefore be included in the portable kernel backend.

\begin{table*}[!t]
    \centering
    \caption{computational characteristics: $N_p$ denotes the number of macroparticles in a local subdomain, $N_m$ the number of retained azimuthal modes, $N_z$ and $N_r$ the longitudinal and radial grid dimensions, respectively, and $S$ the number of grid points covered by the particle shape function. The dominant per step operations can then be grouped into five computational motifs.}
    \label{tab:chara}
    \small
    \setlength{\tabcolsep}{5pt}
    \begin{tabular*}{\textwidth}{@{\extracolsep{\fill}}lcc}
    \toprule
    Component                 & Approximate work &Parallel pattern \\
    \midrule
    Field gather              & $O(N_p N_m S)$ & One or more threads per particle \\
    Particle push             & $O(N_p)$ & Independent particle updates \\
    Charge/current deposition & $O(N_p N_m S)$ & Particle-parallel scattered updates \\
    Fourier-Hankel transforms & $O[N_m(N_r N_z \log N_z + N_z N_r^2)]$ & Mode and grid parallel \\
    PSATD field update        & $O(N_m N_z N_r)$ & Independent spectral-grid updates \\
    \bottomrule
    \end{tabular*}
\end{table*}

\subsection{Kernel execution layer for DCU platforms}
\label{subsec:dcu-kernel-layer}
 DCU accelerators are programmed through a HIP-compatible heterogeneous-computing environment. HIP adopts a CUDA-like host/device model in which a kernel launch is defined by a grid of thread blocks, and the threads within a block can cooperate through synchronization and on-chip shared memory. This similarity reduces the structural effort required to translate CUDA C/C++ kernels, but it does not make Python accelerator backends automatically portable. The original Numba-based GPU path cannot be reused directly in the target environment. Numba officially classified its ROCm target as unmaintained in version 0.54.0 and moved the corresponding implementation outside the main repository. Consequently, the original FBPIC kernels cannot be used directly as a stable and performance-controllable execution path in the DCU software environment.
 
To remove this dependency, the performance-critical kernels are rewritten as explicit C/C++ GPU kernels using the common subset of CUDA and HIP kernel syntax. The kernels retain the grid-block-thread programming model but no longer depend on Python JIT translation. More importantly, it separates kernel implementation from the Python compiler ecosystem and places the computational core closer to the native programming model of both CUDA and HIP devices.

CuPy RawKernel is used as the interface between the rewritten kernels and the FBPIC Python layer. Compilation occurs on first use and is skipped when a valid cached binary is available. On CUDA builds, RawKernel provides NVRTC and NVCC compilation backends, whereas
CuPy's ROCm build provides a HIP-backed implementation of the same kernel
compilation interface\cite{cupy2026rawkernel,cupy2026rocm}. This allows the Python-side kernel invocation mechanism to be retained across CUDA and ROCm builds. Thus, a kernel written within the CUDA/HIP-compatible language subset can be compiled through either the CUDA or ROCm toolchain without changing the FBPIC Python call site.

\begin{figure*}[!t]
    \centering
    \includegraphics[width=\textwidth,trim=5.3bp 5.3bp 5.3bp 5.3bp,clip]{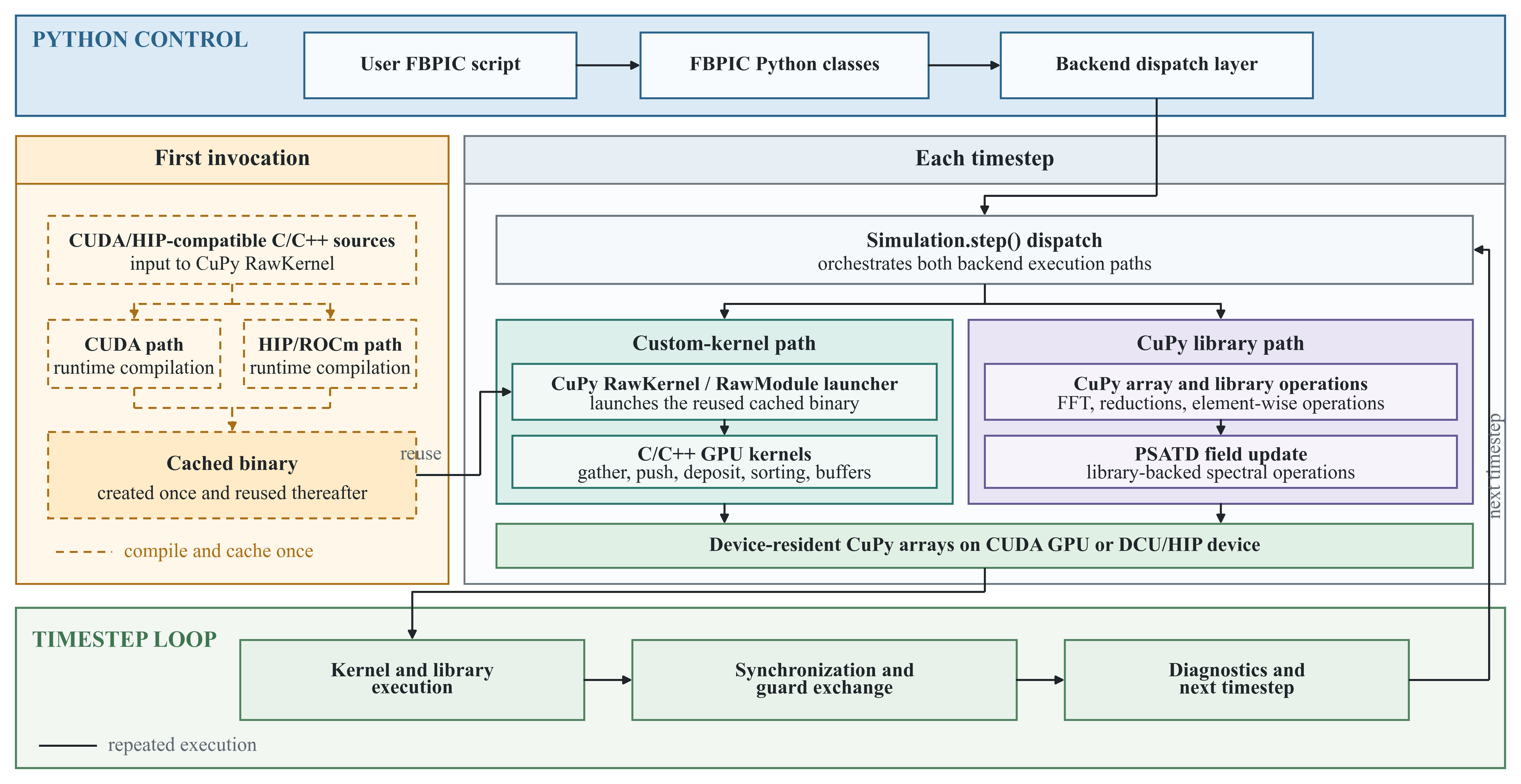}
    \caption{CuPy based accelerator backend of the ported FBPIC implementation. Dashed boxes and arrows denote runtime compilation and binary caching performed upon the first invocation, whereas solid arrows indicate operations repeated at each time step. The cached binary is reused by the custom-kernel path, and both paths operate on device-resident CuPy arrays on a CUDA GPU or DCU/HIP device. Each time step concludes with synchronization and guard exchange, followed by diagnostics, before control returns to Simulation for the next time step.}
    \label{fig:backend-architecture}
\end{figure*}

Figure~\ref{fig:backend-architecture} presents the software architecture of the ported FBPIC implementation. The architecture consists of three layers: the unchanged Python application layer, a backend integration layer, and the device-kernel layer.

Python application and simulation-control layer. The simulation continues to be initiated by a standard FBPIC input script. Users define the computational domain, particle species, laser parameters, boundary conditions, diagnostics, and time-advancement settings through the original Python interface. The Simulation, Particles, Fields, and BoundaryCommunicator objects retain their roles in data ownership and PIC-cycle scheduling. Existing physical models and input scripts therefore require little or no modification.
 
Backend dispatch and kernel-construction layer. The original Numba launch sites are redirected to a CuPy-based backend. This layer selects the appropriate kernel variant, prepares scalar parameters, determines the grid and thread-block configuration, and constructs or retrieves the corresponding RawKernel object. Platform-dependent compiler options and kernel specializations are confined to this layer. The high-level particle and field modules do not need to distinguish between a CUDA GPU and a DCU/HIP device.

Device-data and kernel layer. Particle and field state remains represented by CuPy arrays throughout the simulation. The same device allocation is therefore visible to CuPy array operations, FFT routines, and RawKernel launches, avoiding redundant device copies and preserving compatibility with the existing FBPIC storage model.

The architecture follows three design principles. First, the user-facing Python interface and the existing physical data model are preserved. Second, particle and field data remain device-resident throughout the PIC loop, minimizing host-device traffic. Third, common kernel source is shared between CUDA and HIP wherever practical, while architecture-specific variants are permitted when required for correctness or performance. Together, these principles provide a maintainable path for extending FBPIC to DCU accelerators without sacrificing its Python-based usability or preventing low-level kernel optimization.

\section{Lessons from the DCU port}
\label{sec4}
Porting FBPIC to the DCU platform while achieving functional portability is insufficient to fully unleash the performance of the new accelerator architecture; therefore, this paper explores the following three aspects on the DCU platform.

\subsection{Portable kernel reimplementation and compiler optimization}
\label{subsec:portable-kernel-optimization}
The performance-critical Numba-CUDA kernels were systematically reimplemented as C/C++ GPU kernels using a restricted programming subset accepted by both the CUDA and HIP compilation environments. The rewritten kernel exposes information to the compiler that is difficult to express through high-level array abstractions. Device-array accesses are replaced by explicit pointer arguments, with read-only inputs declared as const T* and non-aliasing arguments marked with $\_\_restrict\_\_$ where the non-aliasing assumption is valid. Restrict-qualified pointers expose non-aliasing information to the compiler, enabling more aggressive code reordering, common-subexpression elimination, and reuse of loaded values\cite{nvidia2026cuda,amd2026hipextensions}. Small device helpers are inlined selectively, and thread indices, grid dimensions, and kernel arguments are made explicit. These changes can reduce redundant address calculations and enable more aggressive load reuse and common-subexpression elimination. They also make generated instructions, register usage, memory transactions, and spill behavior more directly accessible to compiler reports and profiling tools\cite{ryoo2008carving}.

The original FBPIC GPU backend first compiles Python-defined kernels into PTX using Numba and subsequently loads and launches the generated PTX through CuPy\cite{fbpic2026cuda}. Compared with launching kernels directly through the Numba runtime, this design reduces host-side launch overhead by exploiting CuPy’s lower-overhead execution interface, type-based kernel caching, and preprocessing of array metadata before kernel invocation. In the proposed implementation, the performance-critical kernels are instead expressed directly as C/C++ GPU source code and invoked through CuPy’s runtime-compilation interface. On CUDA platforms, the kernel source is compiled into device code through the CUDA runtime compilation toolchain, cached for subsequent invocations, and launched by directly passing the underlying pointers of CuPy arrays together with scalar arguments. On HIP-based platforms, the Python-side invocation interface remains unchanged, while CuPy maps the kernel compilation and runtime operations to the HIP backend, which generates device-specific code for hip-compatible or DCU accelerators. This unified execution path eliminates the intermediate Python-to-PTX translation stage and reduces the amount of framework-level processing required during kernel preparation and dispatch. Measurements on the evaluated platforms show that the revised backend further decreases first-use compilation latency and host-side kernel-launch overhead, while preserving a consistent high-level programming interface across CUDA and HIP environments.

\subsection{Occupancy optimization}
\label{subsec:occupancy-optimization}
Most computational kernels in FBPIC exhibit substantial thread level parallelism; however, their practical performance is jointly influenced by thread block size, per-thread register usage, memory access patterns, and the overhead of atomic operations\cite{hong2009gpu,amd2026hipperformance,fanfarillo2023registerpressure}. Accordingly, this section investigates occupancy-oriented execution configurations on the DCU platform by accounting for the distinct computational and memory access characteristics of different kernel classes.

\subsubsection{Register pressure}
\label{subsubsec1}
Register pressure is a major determinant of the achievable occupancy of particle kernels. In FBPIC, a particle-processing thread may simultaneously retain the particle position, momentum, charge or macroparticle weight, interpolation coefficients, multiple electromagnetic-field components, and intermediate quantities associated with coordinate transformations or particle updates. The live working set is particularly large in field-gather and current-deposition kernels, where several interpolation weights and accumulation variables may remain active over an extended instruction sequence.

Registers are allocated from finite vector and scalar register files shared by the wavefronts resident on a compute unit. An increase in per-thread register usage therefore reduces the number of wavefronts that can be scheduled concurrently and may weaken the ability of the hardware to hide memory and arithmetic latency. If the compiler cannot accommodate the required live variables in the available register file, some values may be spilled to scratch memory. Because scratch memory is backed by the device memory hierarchy rather than by the register file, spilling introduces additional load and store instructions and can substantially increase kernel latency\cite{fanfarillo2023registerpressure,amd2026hipperformance}. Loop unrolling illustrates the resulting trade-off between instruction efficiency and resource consumption. The interpolation stencil used by the cubic gather operation has a fixed extent known at compilation time. Selectively unrolling such short loops can eliminate loop-control instructions, reduce repeated index calculations, and expose additional instruction-level parallelism to the compiler. However, aggressive unrolling may increase the number and live ranges of intermediate values, thereby increasing register pressure and potentially causing reduced occupancy or spilling\cite{fanfarillo2023registerpressure,amd2026hipperformance}.

For this reason, the DCU implementation employs selective unrolling. Small, fixed-trip-count loops on frequently executed paths are unrolled only when the resulting register allocation remains below an occupancy-relevant threshold. Loops associated with a larger number of intermediate quantities, more complex control flow, or infrequently executed paths retain their iterative form. Variable scopes and temporary values are also restricted where possible to shorten live ranges and allow the compiler to reuse registers. The purpose of these transformations is not simply to minimize the static instruction count, but to obtain a more favorable balance between instruction-level parallelism and resident wavefront parallelism.

\subsubsection{Thread-block configuration optimization}
\label{subsubsec2}
The thread-block configuration determines how GPU threads are organized and therefore affects wavefront utilization, memory-access coalescing, resource allocation, and latency hiding\cite{hong2009gpu,hu2021threadblock,pereira2020block,lurati2024hip}. On the target DCU architecture, threads are scheduled in wavefronts of 64 work-items. Block sizes that are multiples of 64 are therefore generally preferable because they avoid partially occupied wavefronts\cite{liu2024dcu}. Nevertheless, increasing the number of threads per block does not necessarily improve performance. Large blocks may increase register and shared-memory consumption, reduce the number of concurrently resident blocks, and consequently limit the occupancy available for hiding memory and instruction latencies. Moreover, for multidimensional kernels, the block shape influences the mapping between threads and array dimensions and thus affects address continuity, memory coalescing, and atomic-update contention.

The tuning experiments were conducted using a relatively large-scale LWFA simulation. With the original FBPIC thread configuration, the simulation required approximately 150 ms per time step. This workload was selected because it provides sufficient particles and grid points to expose a large number of thread blocks, allowing the major kernels to operate in a throughput-oriented regime rather than being dominated by kernel-launch latency or insufficient parallelism. It also exercises the principal computational stages of FBPIC, including field gathering, particle pushing, charge and current deposition, data-layout transformation, and spectral field operations. The selected workload therefore provides a throughput-oriented test case for evaluating kernel-specific launch configurations under the target DCU architecture and software environment.

\begin{figure*}[!t]
    \centering
    \includegraphics[width=\textwidth,trim=5.3bp 6.9bp 3.3bp 2.4bp,clip]{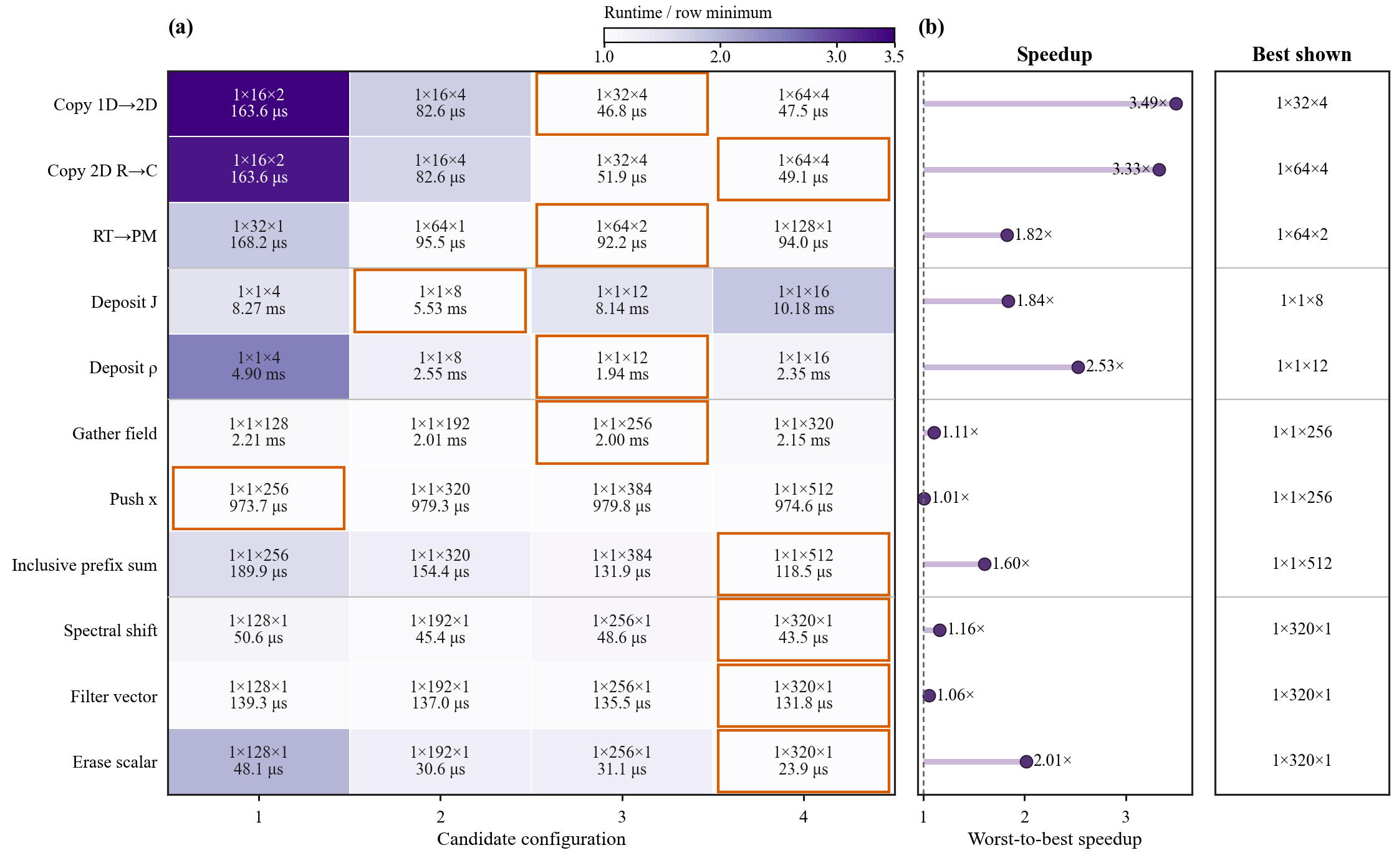}
    \caption{ Kernel-specific thread-block tuning results for FBPIC on the DCU platform using a large-scale LWFA workload. (a) Normalized execution times of candidate block configurations for representative kernels. The color scale denotes the runtime relative to the minimum runtime of each kernel, and orange boxes indicate the selected configurations. Candidates are ordered according to the total number of threads per block and then by block shape. (b) worst-to-best speedup and the selected block configuration for each kernel.}
    \label{fig:thread-block-tuning}
\end{figure*}

For each kernel, a set of candidate configurations was constructed according to its indexing dimensionality and computational structure. One-dimensional particle and grid kernels were evaluated primarily using block sizes ranging from 128 to 512 threads, whereas multidimensional copy, transformation, and deposition kernels were additionally tested with different block shapes. All candidates preserved the numerical algorithm, data layout, and physical simulation parameters; only the kernel-launch geometry was modified. To exclude runtime compilation and first-launch overheads, each configuration was warmed up before measurement and repeatedly executed under identical problem sizes, numerical precision, and compiler settings. The execution time of each candidate configuration was measured using the DCU profiling tool, and the results were normalized to the best-performing configuration for each kernel.
Let \(T_{k,c}\) denote the execution time of kernel \(k\) using
candidate configuration \(c\). The normalized runtime shown in
Figure~\ref{fig:thread-block-tuning}(a) is defined as
\begin{equation}
R_{k,c}
=
\frac{T_{k,c}}{\min_{c'} T_{k,c'}}.
\label{eq:normalized-runtime}
\end{equation}

where a value of \(S_k\) represents the best-performing configuration for a given kernel. The tuning benefit shown in Figure~\ref{fig:thread-block-tuning}(b) is quantified as
\begin{equation}
S_k = \frac{\max_c T_{k,c}}{\min_c T_{k,c}}.
\label{eq:tuning-benefit}
\end{equation}
represents the runtime ratio between the original and best candidate configurations.

As shown in Figure~\ref{fig:thread-block-tuning}, sensitivity to thread-block configuration varies considerably among kernels. Across the 11 representative kernels, the runtime ratio between the worst and best configurations ranges from 1.01 to 3.49. Memory-layout transformation and deposition kernels exhibit the highest sensitivity, whereas several particle and vector kernels maintain near-optimal performance over a relatively broad range of block sizes. 
The application-level benefit of the selected configurations was evaluated using the complete LWFA time-stepping loop. Relative to the launch configurations used by the original FBPIC implementation, the combined kernel-specific configuration reduced the average time per simulation step from 149.171 ms to 112.918 ms. This corresponds to a time reduction of 24.30\%, an overall time-step speedup of 1.32×.

\subsection{Particle locality}
\label{subsec:Particle locality}
Particle reordering is widely used in accelerator-oriented PIC implementations to restore spatial locality. By grouping particles that occupy the same or neighboring cells, subsequent particle kernels operate on spatially coherent subsets of the particle distribution. This improves the locality of field gathering. In WarpX, periodic particle sorting was shown to substantially improve cache reuse and the performance of field gathering and current deposition on GPUs\cite{myers2021warpx}.However, particle sorting itself introduces additional work. A conventional sorting procedure requires the spatial key of every particle to be generated, a permutation to be constructed, and multiple particle attribute arrays to be physically rearranged. Because particle positions change continuously, this organization must be periodically reconstructed. For simulations with a large number of macroparticles or species carrying many auxiliary attributes, the associated memory traffic can become non-negligible and partially offset the performance gained from improved locality.

We examine the particle organization strategy used by FBPIC and develop a lightweight binning sorting alternative that avoids physically reordering all particle attributes. The following subsections first discuss the deposition algorithm, which motivates the need for spatial particle organization, and then compare the original sorting procedure with the proposed binning/counting approach.
\subsubsection{Current deposition}
\label{subsubsec:current-deposition}
Among the major stages of a PIC timestep, particle pushing is comparatively straightforward to parallelize because the state of each particle can largely be updated independently. Field gathering is also naturally particle parallel: each execution thread reads grid values surrounding a particle and writes the interpolated fields only to that particle. Current and charge deposition, in contrast, perform the reverse particle-to-grid operation and are therefore more challenging on massively parallel architectures. Contributions from multiple particles may overlap on the same grid locations, creating concurrent updates that must be handled without introducing data races\cite{stantchev2008particlegrid,kong2011gpupic,myers2021warpx}.

A straightforward particle-parallel implementation can resolve these conflicts using global-memory atomic operations. Although atomics preserve correctness, their performance depends strongly on the spatial distribution of particles. When many concurrently processed particles contribute to overlapping grid points, the corresponding memory updates contend for the same locations, limiting the amount of effective parallelism. Alternatively, private or shared-memory accumulation buffers can reduce the frequency of global atomic updates, but their storage requirements increase rapidly with the grid stencil, number of field components, and number of particles processed concurrently. Excessive use of such buffers can increase shared-memory or register pressure and consequently reduce kernel occupancy.

FBPIC adopts a different organization that exploits particle locality at the cell level. Particles are first classified according to the cell involved in their deposition stencil. The resulting cell indices are ordered, and a prefix-sum array records the range of particles belonging to each cell. The GPU deposition kernel can therefore assign work at the cell level rather than treating all particle contributions as independent global-memory updates. Contributions from the particles associated with a cell are first accumulated into thread-local variables, after which the accumulated quantities are written to the surrounding grid points. Atomic operations remain necessary because the deposition stencils of neighboring cells overlap, but a potentially large number of particle contributions can be locally aggregated before the corresponding global updates are issued. Particles located within the same cell reuse closely related grid data and share the same deposition neighborhood, allowing their contributions to be processed in a structured manner.

\subsubsection{Particle sorting and binning}
\label{subsubsec:particle-sorting-binning}
Binning and cell-based particle organization are commonly used in GPU PIC implementations to trade sorting overhead for improved locality\cite{kong2011gpupic,myers2021warpx}. We investigated a binning/counting strategy for the DCU backend. Instead of constructing a complete ordering and physically rearranging every particle attribute, particles are first mapped to spatial bins. A bin may correspond to an individual grid cell or, more generally, to a super-cell composed of several neighboring cells. A counting kernel determines the number of particles assigned to each bin, and an exclusive prefix scan converts these counts into offsets in an auxiliary index space. Particle indices are then placed into their corresponding bin ranges. Subsequent kernels can traverse particles according to these ranges while leaving the original particle attribute arrays unchanged.

The principal advantage of this approach is that the volume of data physically moved during particle organization is substantially reduced. The method reorganizes a compact array of particle indices rather than repeatedly permuting all position, momentum, weight, and auxiliary particle arrays. Its complexity is also closer to linear in the number of particles and bins, consisting primarily of particle classification, histogram construction, prefix scan, and index generation. The counting stage introduces concurrent updates when multiple particles are assigned to the same bin and therefore generally requires atomic increments or an equivalent privatized histogram scheme. The prefix scan and construction of the binned-index array introduce additional global-memory passes. Moreover, because particle attributes remain in their original locations, processing particles through an index array introduces a level of indirection that can weaken the memory-access regularity of particle attributes compared with a fully reordered structure.

Our measurements show that, despite reducing the amount of full-array particle movement, the binning/counting implementation results in a total timestep cost comparable to that of the original physical-reordering scheme. This observation indicates that the cost removed from particle-array permutation is largely replaced by histogram construction, prefix processing, index generation, and indirect memory accesses in subsequent kernels. More importantly, it demonstrates that minimizing sorting complexity in isolation does not necessarily minimize end-to-end PIC runtime.

\section{Verification of numerical consistency}
\label{sec5}
Before evaluating performance, we assessed whether replacing the Numba-based GPU kernels with explicit C/C++ kernels changed the numerical results of FBPIC. We used two complementary tests. The linear-wakefield benchmark compares each implementation with an analytical solution, whereas the nonlinear LWFA benchmark evaluates cross-backend consistency during coupled particle-field evolution. Hereafter, the upstream Numba-CUDA implementation is denoted as Original FBPIC, the ported backend running on NVIDIA GPUs as Ported CUDA, and the ported backend running on the DCU/HIP platform as Ported DCU.

\subsection{Linear wakefield benchmark}
\label{subsec:linear-wakefield-benchmark}
We first evaluated the numerical accuracy and cross-backend consistency of the three implementations using the FBPIC linear-wakefield verification test\cite{fbpic2026linearwakefield}. This test exercises the complete PIC cycle by simulating a linear
laser-driven plasma wakefield and comparing the longitudinal and radial electric fields, $E_z$ and $E_r$, with analytical reference solutions derived in the linear wakefield regime
\cite{esarey2009laser}.

The benchmark used \(N_m=2\) azimuthal modes and a linearly polarized Gaussian laser pulse. The domain contained \(N_z=800\) longitudinal and \(N_r=120\) radial grid points, with longitudinal and radial extents of \(40\ \mu\mathrm{m}\) and \(60\ \mu\mathrm{m}\), respectively. The plasma density was \(8\times10^{24}\ \mathrm{m^{-3}}\). The laser had a normalized amplitude \(a_0=0.01\), a waist of \(20\ \mu\mathrm{m}\), and a pulse length \(c\tau=6\ \mu\mathrm{m}\). Each simulation was advanced for 1500 time steps using a moving window that propagated at the speed of light. The physical parameters, grid, particle loading, time step, and diagnostics were identical across the three implementations.

Following the acceptance criterion of the official benchmark, we normalized the maximum absolute error by the maximum analytical field amplitude. For \(E_z\), the resulting errors were 7.408\%, 7.409\%, and 7.410\% for Original FBPIC, Ported CUDA, and Ported DCU, respectively, below the prescribed 8\% limit. The corresponding \(E_r\) errors were 6.508\%, 6.509\%, and 6.507\%, below the 11\% limit. Thus, all three implementations satisfied the analytical verification criteria.

We separately quantified implementation-level differences using the relative \(L_2\) metric

\begin{equation}
\epsilon_2(u)=\frac{\lVert u-u_{\mathrm{orig}}\rVert_2}{\lVert u_{\mathrm{orig}}\rVert_2},
\label{eq:l2-error}
\end{equation}

where \(u_{\mathrm{orig}}\) denotes the field produced by Original FBPIC. For Ported CUDA, \(\epsilon_2\) was \(5.85\times10^{-6}\) for \(E_z\) and \(1.04\times10^{-4}\) for \(E_r\). For Ported DCU, the corresponding values were \(7.98\times10^{-6}\) and \(1.25\times10^{-4}\). These values are substantially smaller than the differences between the numerical and analytical solutions.
\begin{figure*}[!t]
\centering
\includegraphics[width=\textwidth,trim=9.9bp 14.0bp 6.7bp 15.6bp,clip]{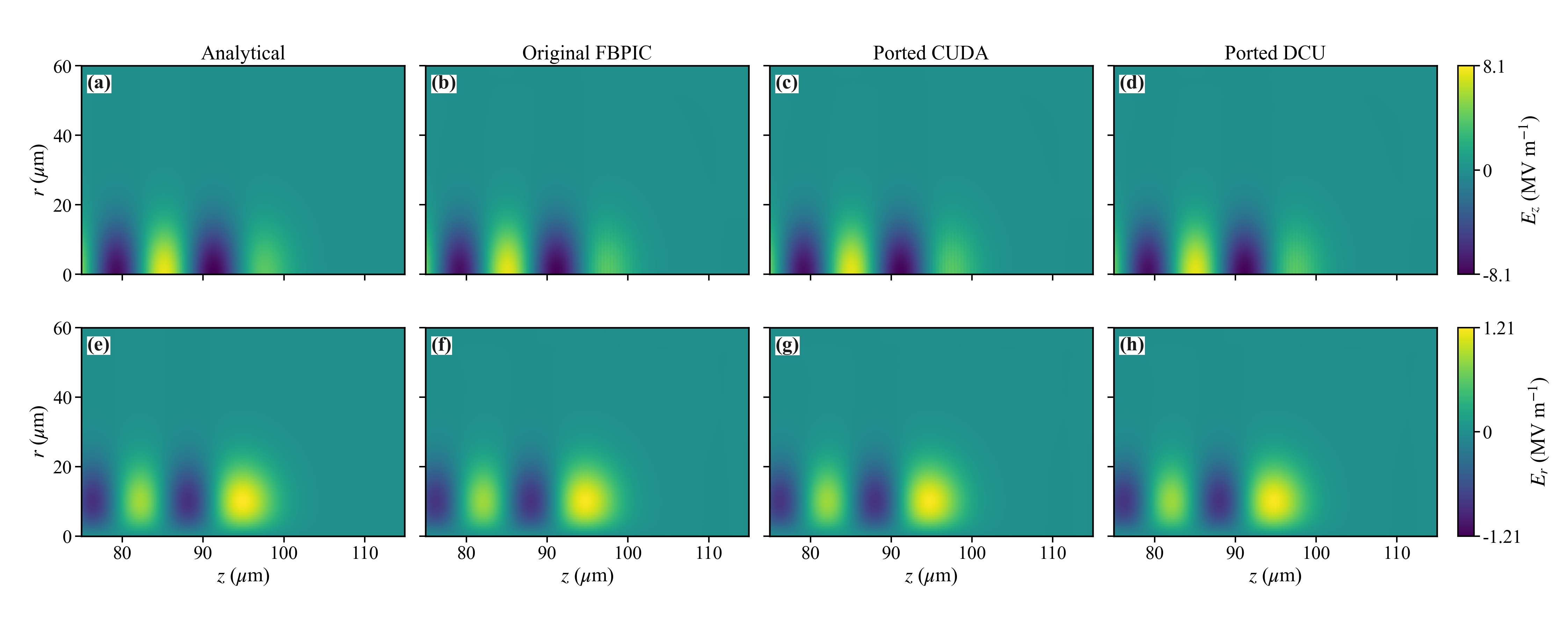}
\caption{Two-dimensional \(r\)-\(z\) distributions of (a-d) the longitudinal electric field \(E_z\) and (e-h) the radial electric field \(E_r\) for the \(N_m=2\) linear-wakefield benchmark after 1500 time steps. From left to right, the columns show the analytical solution, Original FBPIC, Ported CUDA, and Ported DCU. A common symmetric color scale is used within each row.}
\label{fig:linear-wakefield-fields}
\end{figure*}

Figure~\ref{fig:linear-wakefield-fields} compares the two-dimensional distributions of \(E_z\) and \(E_r\). The original CUDA implementation, the ported CUDA implementation, and the DCU implementation reproduce the same longitudinal oscillation, radial field distribution, and spatial localization as the analytical solution. In particular, the positions of the accelerating and decelerating phases in \(E_z\), together with the alternating radial structure of \(E_r\), remain unchanged after the kernel reimplementation. No backend dependent displacement, phase shift, or distortion of the wakefield structure is observed. The field distributions produced by the ported backend on CUDA and DCU are also visually indistinguishable from those obtained with the original FBPIC implementation.
\begin{figure*}[!t]
\centering
\includegraphics[width=\textwidth,trim=15.2bp 16.4bp 3.8bp 6.9bp,clip]{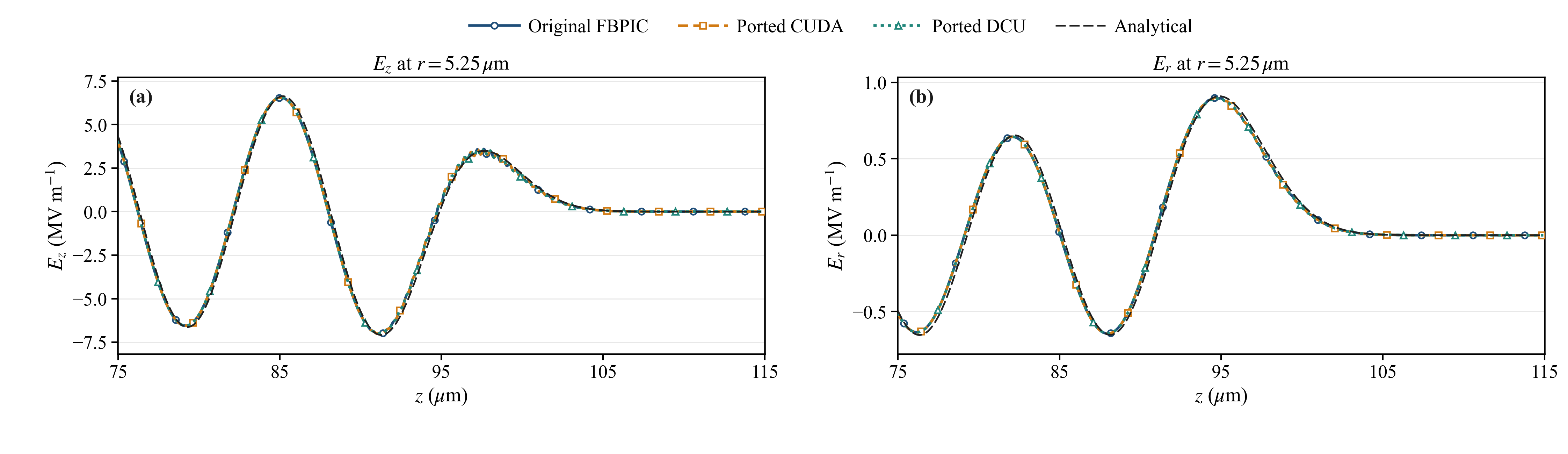}
\caption{Longitudinal lineouts at \(r=5.25\ \mu\mathrm{m}\) for the \(N_m=2\) linear-wakefield benchmark: (a) \(E_z\) and (b) \(E_r\). The curves show Original FBPIC, Ported CUDA, Ported DCU, and the analytical solution.}
\label{fig:linear-wakefield-lineouts}
\end{figure*}
The off-axis lineouts at \(r=5.25\ \mu\mathrm{m}\) in Figure~\ref{fig:linear-wakefield-lineouts} provide a more sensitive comparison. The three numerical curves overlap throughout the displayed longitudinal interval, including the extrema and zero crossings. Their similar deviations from the analytical solution indicate that the backend replacement introduced no additional systematic error pattern at this resolution.

\subsection{LWFA benchmark}
\label{subsec:nonlinear-lwfa-benchmark}
\begin{figure*}[!t]
\centering
\includegraphics[width=0.8\textwidth,trim=6.2bp 14.0bp 9.4bp 15.6bp,clip]{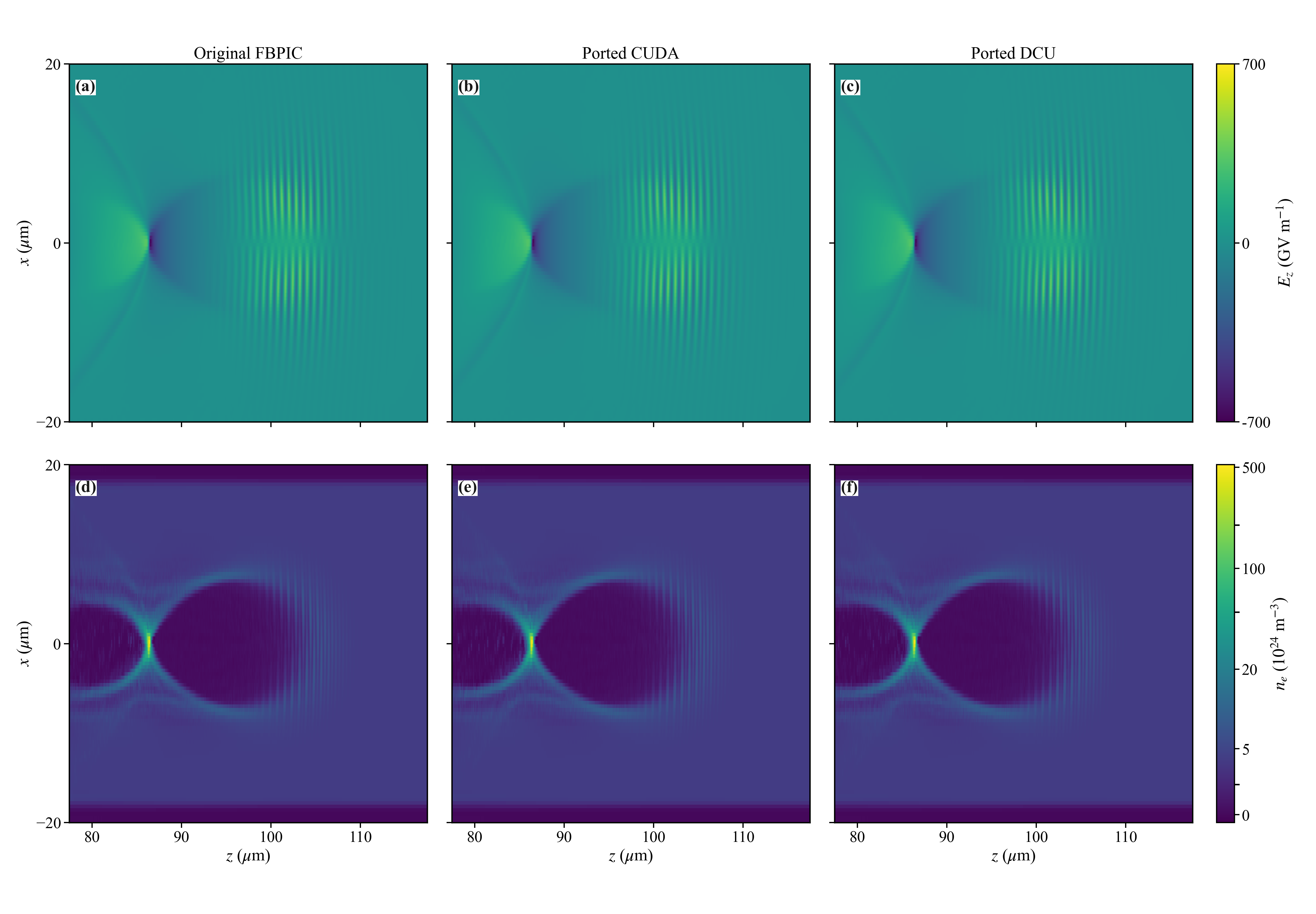}
\caption{Reconstructed \(x\)-\(z\) cross-sections at \(y=0\) for the LWFA benchmark at iteration 1750 (\(t=291.87\ \mathrm{fs}\)). Panels (a-c) show the longitudinal electric field \(E_z\), and panels (d-f) show the electron-density estimator \(n_e=-\rho/e\). The columns correspond to Original FBPIC, Ported CUDA, and Ported DCU. A common color scale is used within each row. The density panels use an asinh normalization to display the background plasma and compressed-density structures on the same scale.}
\label{fig:lwfa-field-density}
\end{figure*}

We further evaluated the numerical consistency of the ported backend using a representative LWFA simulation. Compared with the linear wakefield benchmark, this case involves a substantially higher laser intensity and a stronger plasma response, thereby providing a more representative application level test of the coupled particle and field evolution in FBPIC. The simulation is based on the representative LWFA input example distributed with FBPIC\cite{fbpic2026lwfaexample}.
The computational domain contains \(N_z=800\) longitudinal and \(N_r=50\) radial grid points, with \(N_m=2\) azimuthal modes. The longitudinal simulation window extends from \(-10~\mu\mathrm{m}\) to \(30~\mu\mathrm{m}\), while the radial extent is \(20~\mu\mathrm{m}\). The plasma electron density is \(4\times10^{18}~\mathrm{cm}^{-3}\), with two macroparticles per cell in both the longitudinal and radial directions and four particles in the azimuthal direction. A Gaussian laser pulse with normalized amplitude \(a_0=4\), waist \(w_0=5~\mu\mathrm{m}\), and duration \(\tau=16~\mathrm{fs}\) is used to drive the wakefield. A \(40~\mu\mathrm{m}\) linear density up-ramp is applied at the entrance of the plasma, and the simulation employs a moving window propagating at the speed of light. The interaction length is \(50~\mu\mathrm{m}\). The same physical parameters, discretization, and diagnostic configuration are used for the original FBPIC implementation, the ported implementation on CUDA, and the ported implementation on the DCU/HIP platform.
Figure~\ref{fig:lwfa-field-density} compares the longitudinal electric field \(E_z\) and electron density \(n_e\) obtained from the three implementations at iteration 1750, corresponding to \(t=291.87~\mathrm{fs}\).

As shown in Figure~\ref{fig:lwfa-field-density}(a)-\ref{fig:lwfa-field-density}(c), the three implementations produce nearly identical longitudinal electric-field distributions. Both the position and spatial extent of the wakefield are preserved, including the rapidly oscillating longitudinal field behind the laser pulse and the strong field variation associated with the plasma response. No visible shift in the wakefield phase or change in its spatial morphology is observed between the original FBPIC, ported CUDA, and ported DCU results.

The electron-density distributions in Figure~\ref{fig:lwfa-field-density}(d)-\ref{fig:lwfa-field-density}(f) provide a complementary comparison of the particle dynamics. The laser pulse produces a pronounced electron-depleted region surrounded by a compressed density structure. The location, shape, and extent of this density modulation are consistently reproduced by all three implementations. In particular, the low-density cavity and the high-density electron accumulation near its boundary occur at essentially the same longitudinal and radial positions. Since the electron-density distribution is determined by repeated field gathering, particle pushing, and charge and current deposition over many time steps, its close agreement provides a sensitive application-level indication that the rewritten particle kernels preserve the behavior of the original implementation.

\begin{figure*}[!t]
\centering
\includegraphics[width=0.8\textwidth,trim=6.2bp 14.0bp 7.8bp 13.4bp,clip]{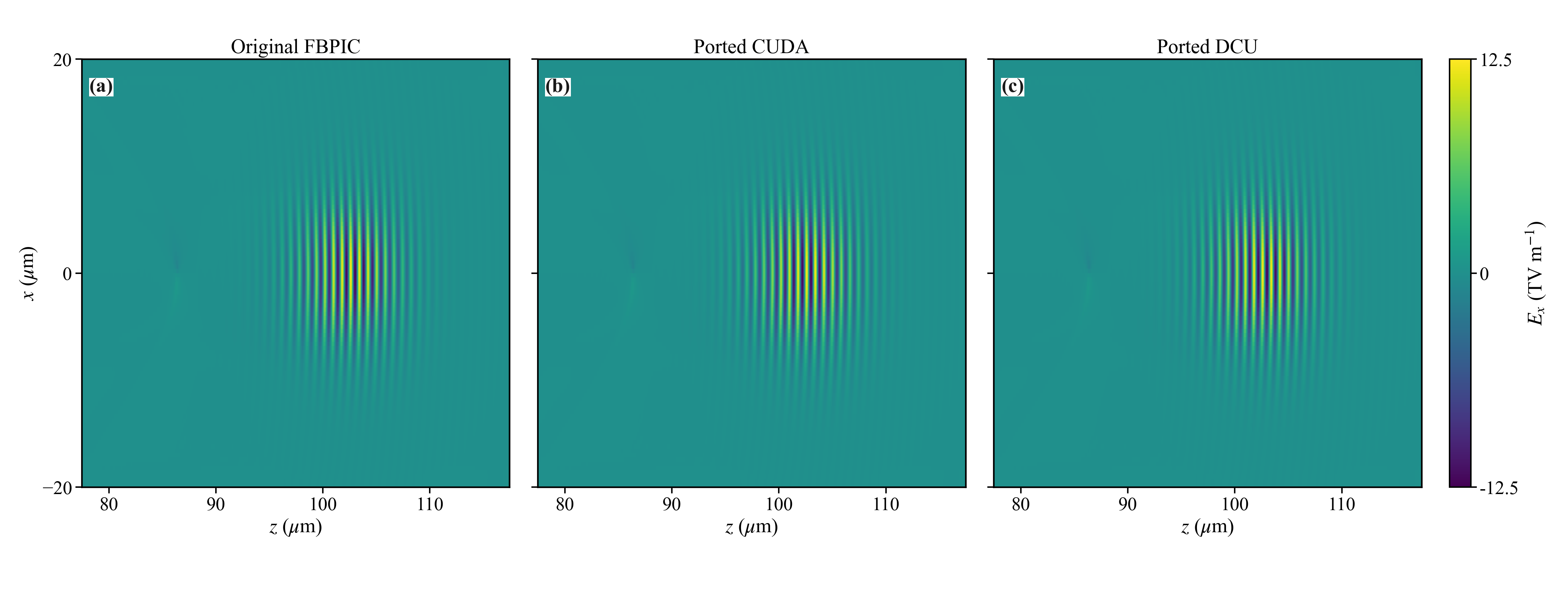}
\caption{Signed transverse Cartesian electric field \(E_x\) in the reconstructed \(x\)-\(z\) plane at \(y=0\) and iteration 1750 (\(t=291.87~\mathrm{fs}\)): (a) Original FBPIC, (b) Ported CUDA, and (c) Ported DCU. A common symmetric color scale is used in all panels.}
\label{fig:lwfa-transverse-laser-field}
\end{figure*}

Figure~\ref{fig:lwfa-transverse-laser-field} compares the signed transverse Cartesian electric field \(E_x\) in the reconstructed \(x\)-\(z\) plane. The spatial region occupied by the laser pulse and its internal oscillation pattern are consistent across the three implementations. The corresponding \(E_x\) values of \(\epsilon_2\) were \(5.19\times10^{-4}\) for Ported CUDA and \(4.86\times10^{-4}\) for Ported DCU. No backend-dependent phase displacement is apparent at the plotted resolution.

\section{Performance results}
\label{sec6}

\subsection{Experimental platform}
\begin{table}[!tbp]
\centering
\caption{Hardware and software configurations used in the performance evaluation.}
\label{tab:experimental-platform}
\footnotesize
\setlength{\tabcolsep}{3.5pt}
\renewcommand{\arraystretch}{1.12}
\begin{tabularx}{\columnwidth}{@{}>{\raggedright\arraybackslash}p{0.33\columnwidth}
                                    >{\raggedright\arraybackslash}X
                                    >{\raggedright\arraybackslash}X@{}}
\toprule
\textbf{Item} & \textbf{NVIDIA CUDA platform} & \textbf{Hygon DCU platform} \\
\midrule
Device
& NVIDIA Tesla V100 PCIe
& Hygon BW1000 DCU \\

Memory
& 32 GB HBM2
& 64 GB HBM2e \\

Peak FP64 performance
& 7.0 TFLOPS
& 30 TFLOPS \\

Peak memory bandwidth
& 900 GB/s
& 1.8 TB/s \\

Software stack
& CUDA Toolkit 12.6
& DTK 25.04 \\
\bottomrule
\end{tabularx}
\end{table}

The hardware platforms used in the performance evaluation are summarized in Table~\ref{tab:experimental-platform}. The CUDA experiments were conducted on an NVIDIA Tesla V100 GPU, whereas the DCU experiments were performed on the Sugon 8000 (Dengfeng) supercomputing system equipped with Hygon BW1000 heterogeneous high-performance computing accelerators. The BW1000 is a general-purpose DCU accelerator designed for high-performance computing and heterogeneous workloads and provides 64~GB of HBM2e memory. At the system level, the Sugon 8000 provides a large-scale heterogeneous computing environment interconnected through a scaleFabric high-speed network with native RDMA capability and supported by the ParaStor distributed storage system. This infrastructure provides the inter-node communication and storage environment used in the multi-DCU scalability experiments. The NVIDIA platform used CUDA Toolkit 12.6, including CUDA compiler version 12.6.20. The DCU platform used DTK 25.04. DTK (DCU Toolkit) is the software stack provided for Hygon DCU accelerators and includes the HIP-compatible programming environment, compiler toolchain, runtime libraries, and accelerator libraries required for heterogeneous computing. On both platforms, the ported FBPIC backend retains the same Python-side execution model based on CuPy, while the rewritten C/C++ kernels are compiled and executed through the corresponding CUDA or DTK/HIP backend.

The NVIDIA V100 platform was used to compare the original FBPIC implementation with the ported CUDA backend, thereby evaluating whether the backend reimplementation affects performance on the original CUDA platform. The Sugon 8000 platform was used for the DCU performance evaluation and multi-accelerator scalability studies described in the following sections.

\subsection{Single accelerator performance}
\label{subsec:single-gpu-performance}
To evaluate the single-accelerator performance of the proposed backend, we employ a LWFA benchmark, which is also used in the subsequent performance and scalability evaluations unless otherwise specified. LWFA represents a typical plasma-acceleration problem involving the interaction between an intense laser pulse and plasma. As the laser propagates through the plasma, it drives plasma-electron oscillations and generates a wakefield that can be exploited to accelerate charged particles. The simulation involves the major computational components of FBPIC, including electromagnetic field evolution, particle pushing, and particle-mesh interpolation and current deposition, and therefore provides a representative workload for assessing both computational and communication performance in realistic plasma-acceleration applications.
For the single-accelerator evaluation,we compare three execution configurations: the original FBPIC implementation on an NVIDIA V100 GPU (Original FBPIC), the ported backend running on the same V100 GPU (Ported CUDA), and the ported backend running on a single DCU accelerator (Ported DCU). The average execution time per simulation step is used as the performance metric.

Three problem sizes are considered by varying the longitudinal and radial grid resolutions while keeping the remaining physical and numerical parameters unchanged. The largest case uses \(N_z=2452\) and \(N_r=512\). The second case reduces the longitudinal resolution to \(N_z=1226\) while retaining \(N_r=512\), and the third case further reduces the radial resolution to \(N_r=256\). These configurations provide workloads with different computational granularities and therefore allow the influence of the rewritten kernel execution path to be examined over a range of problem sizes.

\begin{figure}[!tbp]
    \centering
    \includegraphics[width=\columnwidth,trim=22.6bp 6bp 22.5bp 7bp,clip]{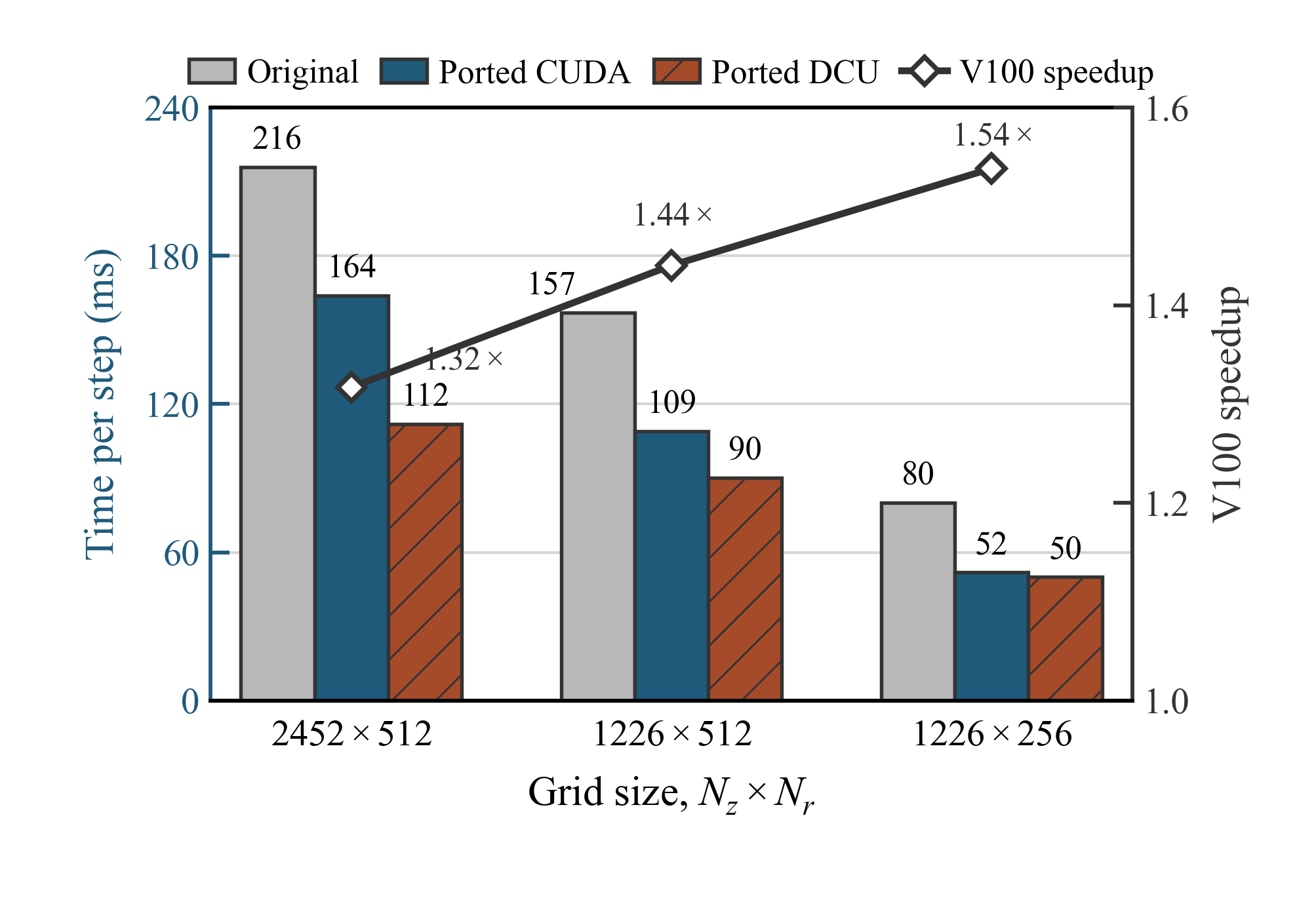}
    \caption{Single-accelerator performance of the LWFA benchmark for three grid configurations. The bars show the average time per simulation step for Original FBPIC and Ported CUDA on an NVIDIA V100 GPU and for Ported DCU on a single DCU accelerator. The line shows the V100 speedup of Ported CUDA relative to Original FBPIC, defined as \(T_{\mathrm{Original}}/T_{\mathrm{Ported\ CUDA}}\). The ported CUDA backend achieves speedups of \(1.32\times\), \(1.44\times\), and \(1.54\times\) for the three problem sizes.}
    \label{fig:v100}
\end{figure}

As shown in Figure~\ref{fig:v100}, the ported CUDA backend consistently outperforms the original FBPIC implementation on the NVIDIA V100. For the \(2452\times512\) case, the average time per step decreases from 216 ms to 164 ms, corresponding to a speedup of \(1.32\times\). For the \(1226\times512\) case, the runtime is reduced from 157 ms to 109 ms, yielding a \(1.44\times\) speedup. The largest relative improvement is observed for the \(1226\times256\) case, for which the runtime decreases from 80 ms to 52 ms and the speedup reaches \(1.54\times\). These reductions correspond to approximately 24.1\%, 30.6\%, and 35.0\% of the original V100 execution time, respectively.
The increasing speedup as the problem size decreases suggests that the revised backend improves not only kernel execution but also the fixed overhead associated with the original Numba-based kernel path. Such overhead represents a larger fraction of the time step for smaller workloads, making the benefit of the direct C/C++ kernel implementation and CuPy-based invocation path more visible.

The Ported DCU configuration requires 112, 90, and 50 ms per step for the three problem sizes, respectively. These results demonstrate that the same ported backend can execute the LWFA workload efficiently on the target HIP-compatible accelerator without changing the high-level FBPIC simulation workflow. Since the V100 and DCU measurements are obtained on different accelerator architectures, their absolute runtimes are reported as a platform-level comparison rather than as an architecture-normalized speedup. Overall, the results show that extending FBPIC to the DCU platform does not compromise its CUDA performance; instead, the rewritten backend improves execution efficiency on the original NVIDIA platform while providing efficient execution on the target DCU system.

\subsection{Strong scaling performance on DCUs}
\label{subsec:strong-scaling}

To characterize the multi-accelerator scalability of the ported backend, we first perform a strong-scaling experiment using the LWFA benchmark. The global problem size is fixed at \(N_z=2453\), \(N_r=512\), and \(N_m=3\), while the number of MPI ranks is increased from 1 to 8. Each MPI rank is mapped to one DCU. A finite-order stencil with \(n_{\mathrm{order}}=32\) is used to enable longitudinal domain decomposition and guard-cell exchange, while all other physical and numerical parameters are kept unchanged.

Each configuration is first executed for 20 warm-up time steps to exclude initialization and runtime-compilation overheads, followed by 500 measured time steps. Device-side CuPy events are used for timing. Because the progress of the distributed simulation is determined by the slowest rank, the maximum of the average step times over all MPI ranks is reported as the parallel execution time. The strong-scaling efficiency for \(N\) DCUs is defined as

\begin{equation}
E_{\mathrm{s}}(N)=\frac{T_1}{N T_N},
\label{eq:strong-scaling-efficiency}
\end{equation}

where \(T_1\) is the average time per step on one DCU and \(T_N\) is the corresponding time using \(N\) DCUs.

\begin{figure}[!tbp]
    \centering
    \includegraphics[width=\columnwidth,trim=22.6bp 19.8bp 22.5bp 4.2bp,clip]{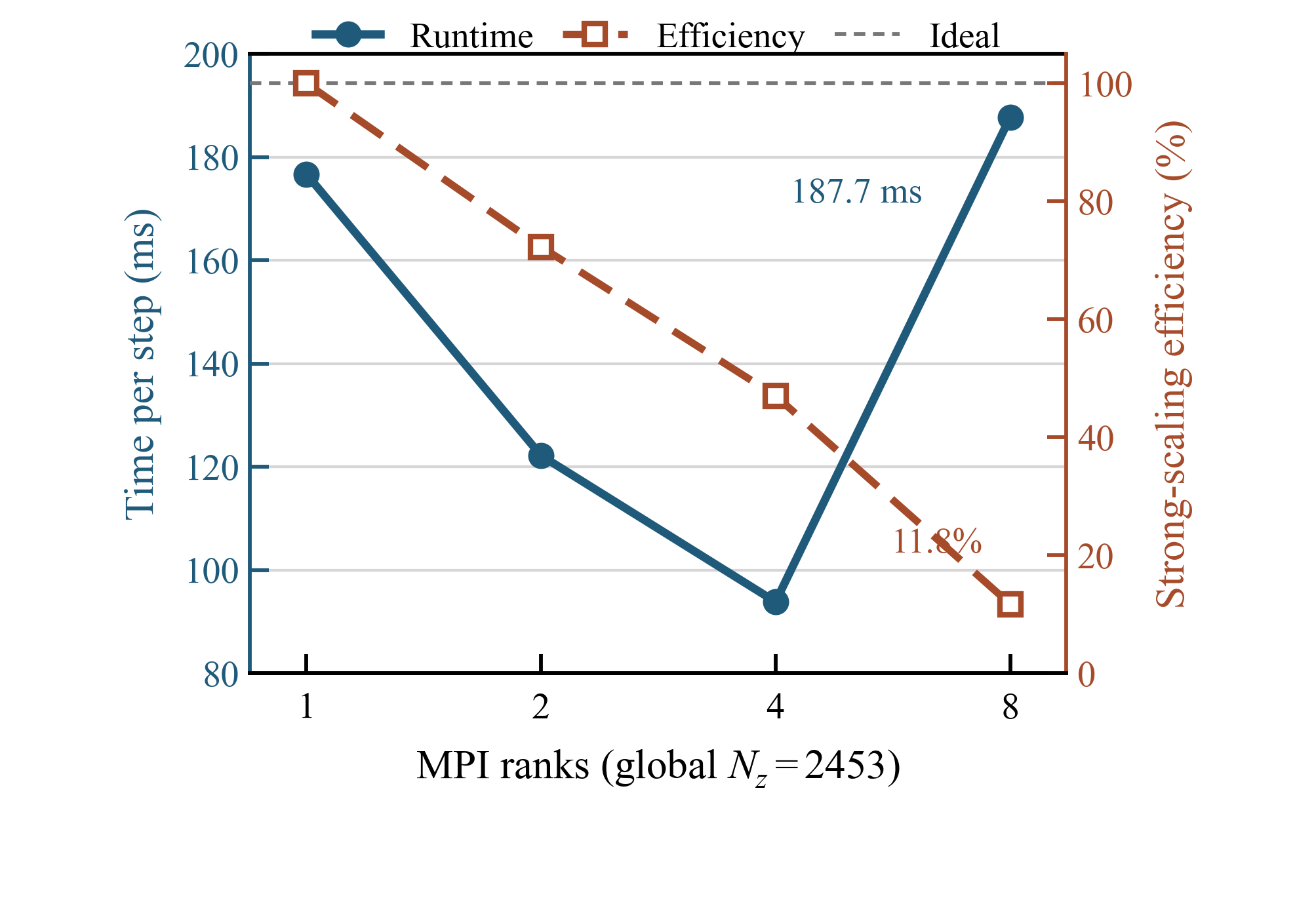}
    \caption{Strong-scaling performance of the ported FBPIC implementation for the LWFA benchmark on the DCU platform. The average time per simulation step and the corresponding strong-scaling efficiency are shown as functions of the number of MPI ranks, with one DCU assigned to each rank. The global problem size is fixed at \(N_z=2453\), \(N_r=512\), and \(N_m=3\). The dashed horizontal line denotes the ideal strong-scaling efficiency of 100\%. The minimum measured runtime is obtained with four DCUs, whereas the eight-DCU configuration exhibits a pronounced loss of efficiency.}
    \label{fig:strong}
\end{figure}

Figure~\ref{fig:strong} shows that increasing the number of DCUs initially reduces the time per step, but the improvement is substantially below ideal strong scaling. The runtime decreases from approximately 177 ms on one DCU to 122 ms on two DCUs, corresponding to a speedup of approximately \(1.45\times\) and a parallel efficiency of about 72\%. With four DCUs, the runtime is further reduced to approximately 94 ms, giving the best measured time-to-solution and a speedup of approximately \(1.88\times\). The corresponding efficiency decreases to approximately 47\%, indicating that parallel overhead is already significant even though additional computational resources still reduce the overall execution time.
Increasing the configuration from four to eight DCUs reverses this trend. The average time per step rises to 187.7 ms, and the strong-scaling efficiency falls to 11.8\%. Thus, the eight-DCU configuration is slower than the four-DCU case and provides no time-to-solution advantage over the single-DCU execution for this problem size. Under strong scaling, the longitudinal extent assigned to each rank decreases as the rank count increases, whereas guard-cell exchange and synchronization are still required at every subdomain boundary. Consequently, the amount of useful computation available to amortize the inter-rank overhead progressively decreases.

An additional change occurs at eight ranks in the evaluated system: the 2- and 4-rank configurations execute within a single compute node, whereas the 8-rank configuration spans two nodes. The sharp degradation at this point therefore suggests an additional contribution from inter-node communication and synchronization. This interpretation is examined quantitatively using the mpiP profiles in Section 6.5. Overall, for the fixed LWFA workload considered here, four DCUs provide the minimum execution time, while further decomposition produces an unfavorable computation-to-communication ratio.

\subsection{Weak scaling performance on DCUs}
\label{subsec3}

The weak-scaling behavior of the ported FBPIC implementation is evaluated using the same LWFA workload. The number of MPI ranks is increased from 1 to 8, with one DCU assigned to each rank. To keep the local longitudinal workload approximately constant, the global grid size \(N_z\) is increased proportionally with the number of ranks, taking values of 1200, 2400, 4800, and 9600 for 1, 2, 4, and 8 ranks, respectively. The longitudinal domain extent is increased accordingly, while the grid spacing and the remaining physical and numerical parameters are kept unchanged. Thus, each MPI rank contains approximately 1200 longitudinal grid points throughout the experiment.

As in the strong-scaling measurements, 20 warm-up steps are performed before timing, followed by 500 measured steps. The maximum average step time among all MPI ranks is used as the parallel runtime. The weak-scaling efficiency is defined as
\begin{equation}
E_{\mathrm{weak}}(p) = \frac{T_1}{T_p},
\label{eq:weak-scaling-efficiency}
\end{equation}

where \(T_1\) and \(T_p\) denote the average time per time step obtained using
one and \(p\) MPI processes, respectively. The corresponding normalized
aggregate throughput is defined as
\begin{equation}
S_{\mathrm{weak}}(p)
= p\frac{T_1}{T_p}
= pE_{\mathrm{weak}}(p).
\label{eq:weak-scaling-throughput}
\end{equation}

\begin{figure}[!tbp]
    \centering
    \includegraphics[width=\columnwidth,trim=22.6bp 12.5bp 22.5bp 4.2bp,clip]{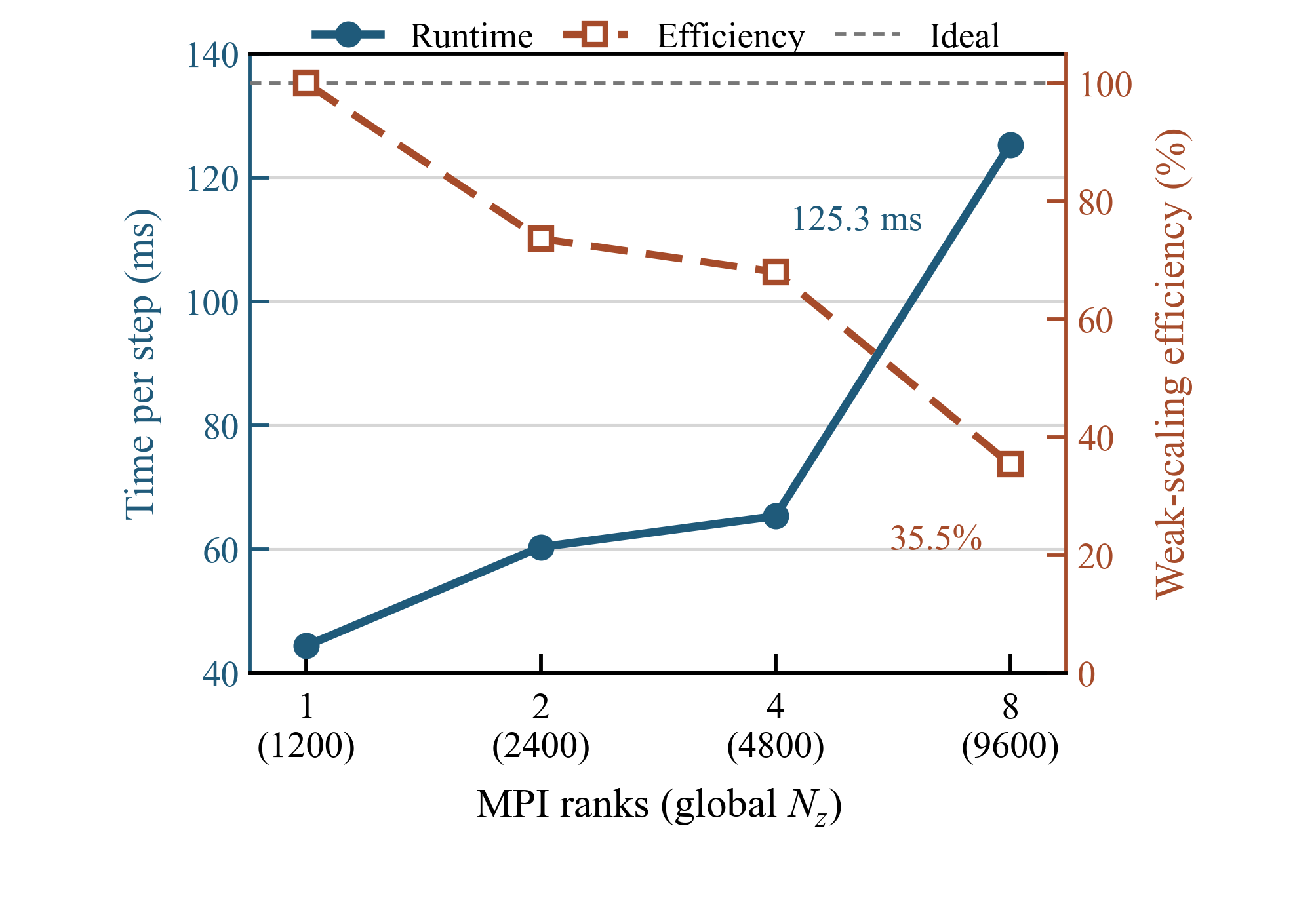}
    \caption{
Weak-scaling performance of the ported FBPIC implementation for the LWFA benchmark on the DCU platform. The average simulation time per step and weak-scaling efficiency are shown for 1, 2, 4, and 8 MPI ranks, with one DCU per rank. The global longitudinal grid size \(N_z\) is increased proportionally from 1200 to 9600 to maintain an approximately constant local workload per rank.
}
    \label{fig:weak}
\end{figure}

Figure~\ref{fig:weak} shows moderate degradation when scaling within a single node. The average time per step increases from approximately 45 ms with one rank to about 60 ms with two ranks, corresponding to a weak-scaling efficiency of approximately 74\%. At four ranks, the runtime increases only moderately further, to approximately 65 ms, while the efficiency remains approximately 68\%. The corresponding normalized aggregate throughputs are approximately \(1.48\times\) and \(2.72\times\) for two and four ranks, respectively. These results indicate that the implementation retains useful weak-scaling behavior up to four DCUs despite the additional boundary communication and synchronization introduced by domain decomposition.

A substantially larger degradation occurs when the execution is extended to eight ranks. The average time per step increases to 125.3 ms and the weak-scaling efficiency decreases to 35.5\%. Although the number of DCUs and the global longitudinal problem size both double from four to eight ranks, the normalized aggregate throughput increases only from approximately \(2.72\times\) to \(2.84\times\), an improvement of only about 4.4\%. Hence, the additional four accelerators contribute little additional aggregate simulation throughput for this configuration.

\subsection{Communication analysis of Multi-DCU scalability}
\label{subsec:mpi-communication-analysis}
To further identify the factors limiting multi-DCU scalability, we profiled the strong- and weak-scaling configurations using \texttt{mpiP}, a lightweight statistical MPI profiling tool\cite{vetter2020mpip}. \texttt{mpiP} defines \texttt{AppTime} as the wall-clock time from the end of \texttt{MPI\_Init} to the beginning of \texttt{MPI\_Finalize}, and MPI time as the wall-clock time accumulated inside MPI calls. Profiles were collected at 2, 4, and 8 ranks for both scaling modes, with one rank mapped to each DCU. The 2- and 4-rank jobs ran within one node, whereas the 8-rank jobs used two nodes with four ranks per node. Because the single-rank case involves no inter-rank communication, the communication analysis begins at two ranks.

\begin{figure*}[!t]
    \centering
    \includegraphics[width=\textwidth,trim=20.2bp 14.7bp 7.1bp 4.1bp,clip]{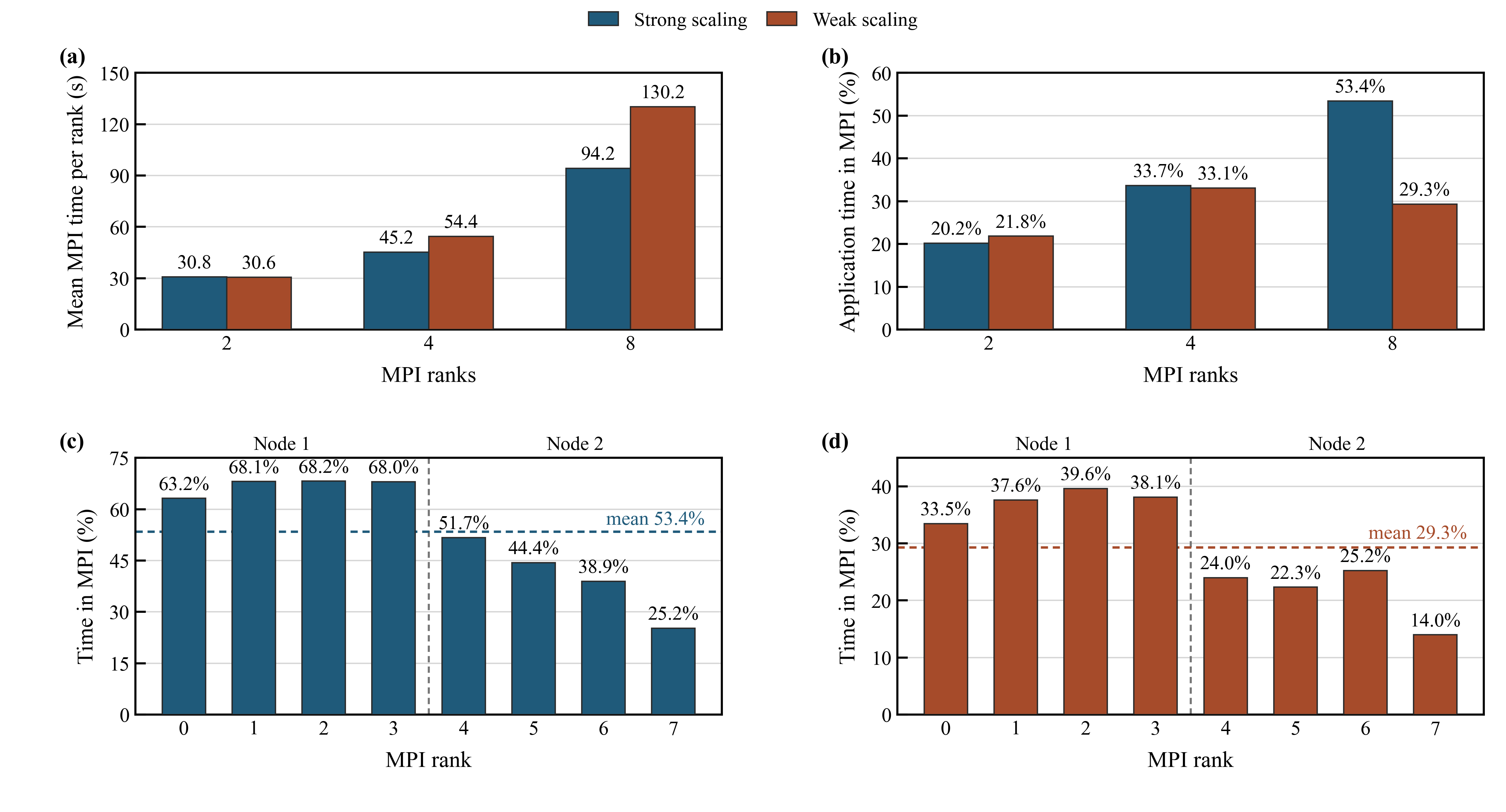}
    \caption{mpiP-based analysis of communication and synchronization overhead in the multi-DCU scaling experiments. (a) Mean cumulative MPI time per rank for the strong- and weak-scaling configurations with 2, 4, and 8 MPI ranks. (b) Fraction of application time spent in MPI routines for the corresponding strong- and weak-scaling configurations. (c) Rank-resolved MPI-time fraction for the 8-rank strong-scaling run. (d) Rank-resolved MPI-time fraction for the 8-rank weak-scaling run. The 2- and 4-rank configurations execute within a single compute node, whereas the 8-rank configuration spans two nodes with four ranks per node. MPI time denotes the wall-clock time accumulated inside MPI routines and may include communication, synchronization, and waiting overhead.}
    \label{fig:mpi-overview}
\end{figure*}

For strong scaling, the increase in MPI-associated overhead closely
correlates with the loss of parallel efficiency. As the rank count
increases from 2 to 4 and 8, the mean cumulative MPI time per rank rises
from 30.8~s to 45.2~s and 94.2~s
(Figure~\ref{fig:mpi-overview}(a)), while the fraction of application time spent
in MPI increases from 20.2\% to 33.7\% and 53.4\%
(Figure~\ref{fig:mpi-overview}(b)). Over the same range, the strong-scaling efficiency decreases from approximately 72\% to 47\% and 11.8\%, as shown in Figure~\ref{fig:strong}. Thus, the computational work removed by domain decomposition is progressively offset by communication and synchronization overhead.

Table~\ref{tab:mpi-communication-pattern} further shows that this degradation is not caused by increasing message size. The aggregate sent volume grows from 53.2 GB at two ranks to 159.1 GB at four ranks and approximately 371 GB at eight ranks, whereas the volume per neighboring interface remains nearly constant at about 53 GB. Likewise, the number of \texttt{MPI\_Isend} calls is approximately 6300 per interface, and the mean message size remains close to 8.4 MB. This behavior is consistent with FBPIC's one-dimensional longitudinal decomposition: adding ranks increases the number of subdomain interfaces, while the communication cost associated with each interface remains approximately unchanged. Under strong scaling, the local particle and grid workload decreases with increasing rank count, whereas the per-interface communication requirement does not decrease at the same rate. The resulting reduction in the computation-to-communication ratio explains the efficiency loss already observed within a single node.

The transition from four to eight ranks introduces an additional penalty because execution changes from intra-node to inter-node communication. The mean MPI time increases from 45.2~s to 94.2~s in the strong-scaling experiment, coinciding with the increase in the average step time from approximately 94~ms to 187.7~ms. This behavior indicates that the inter-node execution regime introduces substantial additional MPI-associated overhead. Since mpiP measures time spent inside MPI routines, these data do not by themselves separate network transfer latency from synchronization and waiting effects, but they clearly show that the inter-node transition is associated with the observed performance reversal. Figure~\ref{fig:mpi-overview}(c) and \ref{fig:mpi-overview}(d) further show that the 8-rank configurations exhibit pronounced rank-level differences in the fraction of time spent in MPI. In the strong-scaling run, the MPI-time fraction ranges from 25.2\% to 68.2\%, while in the weak-scaling run it ranges from 14.0\% to 39.6\%. Moreover, the two groups of ranks located on different nodes exhibit systematically different MPI-time fractions. This nonuniformity indicates unequal MPI-associated waiting or progress across ranks and can further amplify synchronization overhead. The available mpiP data, however, are insufficient to distinguish whether the imbalance originates primarily from communication topology, process placement, or differences in computational progress.

The weak-scaling results exhibit a related but less direct trend. Because the local longitudinal workload is kept approximately constant, the computation available to amortize communication does not decrease with rank count. Consequently, weak scaling remains relatively efficient within a single node: the efficiencies at two and four ranks are approximately 74\% and 68\%, with normalized aggregate throughputs of 1.48$\times$ and 2.72$\times$, respectively. Nevertheless, the mean MPI time per rank increases from 30.6~s at two ranks to 54.4~s at four ranks and 130.2~s at eight ranks. When execution extends to two nodes, the weak-scaling efficiency falls to 35.5\%, while the normalized aggregate throughput increases only from 2.72$\times$ to 2.84$\times$. Although the MPI-time fraction decreases slightly from 33.1\% at four ranks to 29.3\% at eight ranks, the absolute MPI time increases by approximately 2.4$\times$. The lower percentage therefore does not indicate reduced MPI overhead; rather, the total application time grows even more rapidly.

\begin{table*}[!t]
\centering
\caption{Point-to-point communication characteristics from mpiP profiles for the strong- and weak-scaling experiments. The per-interface communication volume is calculated as the aggregate sent volume divided by \(p-1\), where \(p-1\) denotes the number of internal interfaces in the longitudinal domain decomposition.}\label{tab:mpi-communication-pattern}%
\small
\setlength{\tabcolsep}{5pt}
\begin{tabular*}{\textwidth}{@{\extracolsep\fill}lccccc}
\toprule%
Scaling & \(p\) & Sent & Per interface & \texttt{MPI\_Isend} & Mean size \\
 & & (GB) & (GB) & calls & (MB) \\
\midrule
Strong & 2 & 53.2  & 53.2 & 6,300  & 8.38 \\
Strong & 4 & 159.1 & 53.0 & 18,900 & 8.40 \\
Strong & 8 & 371.0 & 53.0 & 44,100 & 8.40 \\
Weak   & 2 & 53.2  & 53.2 & 6,300  & 8.38 \\
Weak   & 4 & 159.1 & 53.0 & 18,900 & 8.37 \\
Weak   & 8 & 370.7 & 53.0 & 44,100 & 8.37 \\
\bottomrule
\end{tabular*}
\end{table*}

These observations are also consistent with the execution characteristics of FBPIC's domain-decomposed parallelization. Inter-device execution introduces boundary communication and synchronization that are absent in the single-device case. It is therefore most effective when the computational workload assigned to each accelerator is sufficiently large to amortize these costs, or when distributed execution is required because the problem cannot fit within the memory of a single device. In the strong-scaling experiment, increasing the number of DCUs progressively reduces the local workload and therefore moves the simulation toward a regime in which the approximately fixed communication cost per neighboring interface can no longer be effectively amortized. In the weak-scaling experiment, preserving the local problem size prevents this reduction in computational granularity, but communication and synchronization overhead remain exposed and increase substantially once execution extends across nodes. 

\section{Conclusion}
\label{sec7}
This paper presented a HIP-compatible accelerator backend for FBPIC that decouples its performance-critical execution path from Numba CUDA while preserving the existing Python interface, device-resident data flow, and MPI-based domain decomposition. The dominant particle and field operations were reimplemented as explicit C/C++ GPU kernels and integrated through CuPy RawKernel, providing a common execution path for CUDA and HIP-compatible platforms. In the linear-wakefield and nonlinear LWFA benchmarks, the ported implementations reproduced the numerical behavior of the original FBPIC code. The relative $L_2$ differences did not exceed $5.19\times 10^{-4}$ for the reported field comparisons, and no backend-dependent phase shift or structural distortion was observed.

The performance results show that the increased portability does not compromise execution efficiency on the evaluated CUDA platform. On an NVIDIA V100 GPU, the ported backend achieved speedups of $1.32\times$, $1.44\times$, and $1.54\times$ for the three LWFA problem sizes. On the target DCU platform, kernel-specific thread-block tuning reduced the average time per simulation step from 149.171 ms to 112.918 ms, corresponding to a 24.30\% reduction and a $1.32\times$ speedup. The multi-DCU experiments achieved their best time-to-solution on four accelerators, with a strong-scaling speedup of $1.88\times$. Under weak scaling, four DCUs delivered $2.72\times$ the single-DCU aggregate throughput at approximately 68\% efficiency. Scaling to eight DCUs produced limited or negative gains. The mpiP profiles attribute this degradation primarily to inter-node communication, synchronization, and the increasingly unfavorable ratio between local computation and the approximately fixed communication cost per subdomain interface.

Although the implementation and tuning experiments in this study were conducted on NVIDIA V100 and DCU accelerators, the underlying approach is not restricted to these particular devices. Replacing a vendor-dependent Python JIT backend with explicitly implemented kernels, preserving device-resident data, and applying compiler- and occupancy-oriented optimization are broadly applicable to other CUDA- and HIP-compatible accelerators. The proposed design therefore provides both a practical accelerator backend for FBPIC and a reference strategy for porting other Python-based scientific applications whose performance-critical components depend on platform-specific JIT frameworks. Future work will focus on reducing multi-node communication overhead, overlapping communication with computation, improving device-aware MPI data exchange, and introducing architecture- and workload-aware kernel configuration mechanisms to extend efficient FBPIC execution to larger DCU systems.


\section*{Acknowledgements}
This work is supported by the National Key Research and Development Program of China (2024YFB4504103), the Fund of Laboratory for Advanced Computing and Intelligence Engineering (2025-ZZKY-016), and the National Natural Science Foundation of China (12574380). This work is also supported by Jiangsu Province Engineering Research Center of IntelliSense Technology and System.

\section*{Data and Code Availability}
The source code developed in this study is available in the Mendeley Data
repository at \url{https://doi.org/10.17632/r66r2vzcjc.1}.
The repository is currently under embargo and will become publicly
available after the embargo period.


\begin{thebibliography}{44}
\expandafter\ifx\csname natexlab\endcsname\relax\def\natexlab#1{#1}\fi
\providecommand{\url}[1]{\texttt{#1}}
\providecommand{\href}[2]{#2}
\providecommand{\path}[1]{#1}
\providecommand{\DOIprefix}{doi:}
\providecommand{\ArXivprefix}{arXiv:}
\providecommand{\URLprefix}{URL: }
\providecommand{\Pubmedprefix}{pmid:}
\providecommand{\doi}[1]{\href{http://dx.doi.org/#1}{\path{#1}}}
\providecommand{\Pubmed}[1]{\href{pmid:#1}{\path{#1}}}
\providecommand{\bibinfo}[2]{#2}
\ifx\xfnm\relax \def\xfnm[#1]{\unskip,\space#1}\fi
%Type = Book
\bibitem[{Birdsall and Langdon(1991)}]{birdsall1991plasma}
\bibinfo{author}{C.~K. Birdsall}, \bibinfo{author}{A.~B. Langdon}, \bibinfo{title}{Plasma Physics via Computer Simulation}, \bibinfo{publisher}{Institute of Physics Publishing}, \bibinfo{address}{Bristol and Philadelphia}, \bibinfo{year}{1991}.
%Type = Incollection
\bibitem[{Fonseca et~al.(2002)Fonseca, Silva, Tsung, Decyk, Lu, Ren, Mori, Deng, Lee, Katsouleas, and Adam}]{fonseca2002osiris}
\bibinfo{author}{R.~A. Fonseca}, \bibinfo{author}{L.~O. Silva}, \bibinfo{author}{F.~S. Tsung}, \bibinfo{author}{V.~K. Decyk}, \bibinfo{author}{W.~Lu}, \bibinfo{author}{C.~Ren}, \bibinfo{author}{W.~B. Mori}, \bibinfo{author}{S.~Deng}, \bibinfo{author}{S.~Lee}, \bibinfo{author}{T.~Katsouleas}, \bibinfo{author}{J.~C. Adam},
\newblock \bibinfo{title}{{OSIRIS: A Three-Dimensional, Fully Relativistic Particle in Cell Code for Modeling Plasma Based Accelerators}},
\newblock in: \bibinfo{booktitle}{Computational Science---ICCS 2002}, \bibinfo{publisher}{Springer Berlin Heidelberg}, \bibinfo{year}{2002}, pp. \bibinfo{pages}{342--351}.
%Type = Article
\bibitem[{Decyk and Singh(2014)}]{decyk2014architectures}
\bibinfo{author}{V.~K. Decyk}, \bibinfo{author}{T.~V. Singh},
\newblock \bibinfo{title}{{Particle-in-Cell algorithms for emerging computer architectures}},
\newblock \bibinfo{journal}{Computer Physics Communications} \bibinfo{volume}{185} (\bibinfo{year}{2014}) \bibinfo{pages}{708--719}.
%Type = Article
\bibitem[{Vay et~al.(2018)Vay, Almgren, Bell, Ge, Grote, Hogan, Kononenko, Lehe, Myers, Ng, Park, Ryne, Shapoval, Thévenet, and Zhang}]{vay2018warpx}
\bibinfo{author}{J.-L. Vay}, \bibinfo{author}{A.~Almgren}, \bibinfo{author}{J.~Bell}, \bibinfo{author}{L.~Ge}, \bibinfo{author}{D.~Grote}, \bibinfo{author}{M.~Hogan}, \bibinfo{author}{O.~Kononenko}, \bibinfo{author}{R.~Lehe}, \bibinfo{author}{A.~Myers}, \bibinfo{author}{C.~Ng}, \bibinfo{author}{J.~Park}, \bibinfo{author}{R.~Ryne}, \bibinfo{author}{O.~Shapoval}, \bibinfo{author}{M.~Thévenet}, \bibinfo{author}{W.~Zhang},
\newblock \bibinfo{title}{{Warp-X: A new exascale computing platform for beam--plasma simulations}},
\newblock \bibinfo{journal}{Nuclear Instruments and Methods in Physics Research Section A: Accelerators, Spectrometers, Detectors and Associated Equipment} \bibinfo{volume}{909} (\bibinfo{year}{2018}) \bibinfo{pages}{476--479}.
%Type = Article
\bibitem[{Liu et~al.(2024)Liu, Hao, Zhang, Lu, Tian, Yang, Xie, Dai, Yuan, Wang, and Yang}]{liu2024dcu}
\bibinfo{author}{Z.~Liu}, \bibinfo{author}{M.~Hao}, \bibinfo{author}{W.~Zhang}, \bibinfo{author}{G.~Lu}, \bibinfo{author}{X.~Tian}, \bibinfo{author}{S.~Yang}, \bibinfo{author}{M.~Xie}, \bibinfo{author}{J.~Dai}, \bibinfo{author}{C.~Yuan}, \bibinfo{author}{D.~Wang}, \bibinfo{author}{H.~Yang},
\newblock \bibinfo{title}{Optimizing depthwise separable convolution on {DCU}},
\newblock \bibinfo{journal}{CCF Transactions on High Performance Computing} \bibinfo{volume}{6} (\bibinfo{year}{2024}) \bibinfo{pages}{646--664}.
%Type = Misc
\bibitem[{{Advanced Micro Devices, Inc.}(2026)}]{amd2026hip}
\bibinfo{author}{{Advanced Micro Devices, Inc.}}, \bibinfo{title}{{HIP} documentation}, \bibinfo{year}{2026}. \URLprefix \url{https://rocm.docs.amd.com/projects/HIP/en/latest/}, \bibinfo{note}{accessed 2026-09-03}.
%Type = Incollection
\bibitem[{Tsai et~al.(2021)Tsai, Cojean, Ribizel, and Anzt}]{tsai2021ginkgo}
\bibinfo{author}{Y.~M. Tsai}, \bibinfo{author}{T.~Cojean}, \bibinfo{author}{T.~Ribizel}, \bibinfo{author}{H.~Anzt},
\newblock \bibinfo{title}{{Preparing Ginkgo for AMD GPUs---A Testimonial on Porting CUDA Code to HIP}},
\newblock in: \bibinfo{booktitle}{Euro-Par 2020: Parallel Processing Workshops}, \bibinfo{publisher}{Springer International Publishing}, \bibinfo{year}{2021}, pp. \bibinfo{pages}{109--121}.
%Type = Article
\bibitem[{Lehe et~al.(2016)Lehe, Kirchen, Andriyash, Godfrey, and Vay}]{lehe2016fbpic}
\bibinfo{author}{R.~Lehe}, \bibinfo{author}{M.~Kirchen}, \bibinfo{author}{I.~A. Andriyash}, \bibinfo{author}{B.~B. Godfrey}, \bibinfo{author}{J.-L. Vay},
\newblock \bibinfo{title}{A spectral, quasi-cylindrical and dispersion-free {Particle-In-Cell} algorithm},
\newblock \bibinfo{journal}{Computer Physics Communications} \bibinfo{volume}{203} (\bibinfo{year}{2016}) \bibinfo{pages}{66--82}.
%Type = Article
\bibitem[{Godfrey et~al.(2014)Godfrey, Vay, and Haber}]{godfrey2014psatd}
\bibinfo{author}{B.~B. Godfrey}, \bibinfo{author}{J.-L. Vay}, \bibinfo{author}{I.~Haber},
\newblock \bibinfo{title}{{Numerical stability analysis of the pseudo-spectral analytical time-domain PIC algorithm}},
\newblock \bibinfo{journal}{Journal of Computational Physics} \bibinfo{volume}{258} (\bibinfo{year}{2014}) \bibinfo{pages}{689--704}.
%Type = Misc
\bibitem[{{FBPIC contributors}(2026)}]{fbpic2026install}
\bibinfo{author}{{FBPIC contributors}}, \bibinfo{title}{{FBPIC} 0.27.0 documentation: Installation on a local computer}, \bibinfo{year}{2026}. \URLprefix \url{https://fbpic.github.io/install/install_local.html}, \bibinfo{note}{accessed 2026-09-03}.
%Type = Inproceedings
\bibitem[{Lam et~al.(2015)Lam, Pitrou, and Seibert}]{lam2015numba}
\bibinfo{author}{S.~K. Lam}, \bibinfo{author}{A.~Pitrou}, \bibinfo{author}{S.~Seibert},
\newblock \bibinfo{title}{{Numba: a LLVM-based Python JIT compiler}},
\newblock in: \bibinfo{booktitle}{Proceedings of the Second Workshop on the LLVM Compiler Infrastructure in HPC}, \bibinfo{publisher}{ACM}, \bibinfo{year}{2015}, pp. \bibinfo{pages}{1--6}.
%Type = Inproceedings
\bibitem[{Okuta et~al.(2017)Okuta, Unno, Nishino, Hido, and Loomis}]{okuta2017cupy}
\bibinfo{author}{R.~Okuta}, \bibinfo{author}{Y.~Unno}, \bibinfo{author}{D.~Nishino}, \bibinfo{author}{S.~Hido}, \bibinfo{author}{C.~Loomis},
\newblock \bibinfo{title}{{CuPy: A NumPy-Compatible Library for NVIDIA GPU Calculations}},
\newblock in: \bibinfo{booktitle}{Proceedings of Workshop on Machine Learning Systems (LearningSys) in the Thirty-first Annual Conference on Neural Information Processing Systems (NIPS)}, \bibinfo{year}{2017}.
%Type = Misc
\bibitem[{{Numba Development Team}(2021)}]{numba2021rocm}
\bibinfo{author}{{Numba Development Team}}, \bibinfo{title}{{Deprecation Notices: ROCm target}}, \bibinfo{howpublished}{Numba 0.54.1 documentation}, \bibinfo{year}{2021}. \URLprefix \url{https://numba.readthedocs.io/en/0.54.1/reference/deprecation.html}, \bibinfo{note}{accessed 2026-07-20}.
%Type = Misc
\bibitem[{{CuPy Development Team}(2026{\natexlab{a}})}]{cupy2026rawkernel}
\bibinfo{author}{{CuPy Development Team}}, \bibinfo{title}{{CuPy} 14.2.0 documentation: cupy.rawkernel}, \bibinfo{year}{2026}{\natexlab{a}}. \URLprefix \url{https://docs.cupy.dev/en/v14.2.0/reference/generated/cupy.RawKernel.html}, \bibinfo{note}{accessed 2026-09-03}.
%Type = Misc
\bibitem[{{CuPy Development Team}(2026{\natexlab{b}})}]{cupy2026rocm}
\bibinfo{author}{{CuPy Development Team}}, \bibinfo{title}{{CuPy} 14.2.0 documentation: Using {CuPy} on {AMD GPU} ({ROCm})}, \bibinfo{year}{2026}{\natexlab{b}}. \URLprefix \url{https://docs.cupy.dev/en/stable/install.html}, \bibinfo{note}{accessed 2026-09-03}.
%Type = Article
\bibitem[{Godfrey(1974)}]{godfrey1974cherenkov}
\bibinfo{author}{B.~B. Godfrey},
\newblock \bibinfo{title}{{Numerical Cherenkov instabilities in electromagnetic particle codes}},
\newblock \bibinfo{journal}{Journal of Computational Physics} \bibinfo{volume}{15} (\bibinfo{year}{1974}) \bibinfo{pages}{504--521}.
%Type = Article
\bibitem[{Vay et~al.(2011)Vay, Geddes, Cormier-Michel, and Grote}]{vay2011mitigation}
\bibinfo{author}{J.-L. Vay}, \bibinfo{author}{C.~Geddes}, \bibinfo{author}{E.~Cormier-Michel}, \bibinfo{author}{D.~Grote},
\newblock \bibinfo{title}{{Numerical methods for instability mitigation in the modeling of laser wakefield accelerators in a Lorentz-boosted frame}},
\newblock \bibinfo{journal}{Journal of Computational Physics} \bibinfo{volume}{230} (\bibinfo{year}{2011}) \bibinfo{pages}{5908--5929}.
%Type = Article
\bibitem[{Myers et~al.(2021)Myers, Almgren, Amorim, Bell, Fedeli, Ge, Gott, Grote, Hogan, Huebl, Jambunathan, Lehe, Ng, Rowan, Shapoval, Th{\'e}venet, Vay, Vincenti, Yang, Za{\"\i}m, Zhang, Zhao, and Zoni}]{myers2021warpx}
\bibinfo{author}{A.~Myers}, \bibinfo{author}{A.~Almgren}, \bibinfo{author}{L.~D. Amorim}, \bibinfo{author}{J.~Bell}, \bibinfo{author}{L.~Fedeli}, \bibinfo{author}{L.~Ge}, \bibinfo{author}{K.~Gott}, \bibinfo{author}{D.~P. Grote}, \bibinfo{author}{M.~Hogan}, \bibinfo{author}{A.~Huebl}, \bibinfo{author}{R.~Jambunathan}, \bibinfo{author}{R.~Lehe}, \bibinfo{author}{C.~Ng}, \bibinfo{author}{M.~Rowan}, \bibinfo{author}{O.~Shapoval}, \bibinfo{author}{M.~Th{\'e}venet}, \bibinfo{author}{J.-L. Vay}, \bibinfo{author}{H.~Vincenti}, \bibinfo{author}{E.~Yang}, \bibinfo{author}{N.~Za{\"\i}m}, \bibinfo{author}{W.~Zhang}, \bibinfo{author}{Y.~Zhao}, \bibinfo{author}{E.~Zoni},
\newblock \bibinfo{title}{Porting {WarpX} to {GPU}-accelerated platforms},
\newblock \bibinfo{journal}{Parallel Computing} \bibinfo{volume}{108} (\bibinfo{year}{2021}) \bibinfo{pages}{102833}.
%Type = Inproceedings
\bibitem[{Fedeli et~al.(2022)Fedeli, Huebl, Boillod-Cerneux, Clark, Gott, Hillairet, Jaure, Leblanc, Lehe, Myers, Piechurski, Sato, Za{\"\i}m, Zhang, Vay, and Vincenti}]{fedeli2022warpx}
\bibinfo{author}{L.~Fedeli}, \bibinfo{author}{A.~Huebl}, \bibinfo{author}{F.~Boillod-Cerneux}, \bibinfo{author}{T.~Clark}, \bibinfo{author}{K.~Gott}, \bibinfo{author}{C.~Hillairet}, \bibinfo{author}{S.~Jaure}, \bibinfo{author}{A.~Leblanc}, \bibinfo{author}{R.~Lehe}, \bibinfo{author}{A.~Myers}, \bibinfo{author}{C.~Piechurski}, \bibinfo{author}{M.~Sato}, \bibinfo{author}{N.~Za{\"\i}m}, \bibinfo{author}{W.~Zhang}, \bibinfo{author}{J.-L. Vay}, \bibinfo{author}{H.~Vincenti},
\newblock \bibinfo{title}{Pushing the frontier in the design of laser-based electron accelerators with groundbreaking mesh-refined particle-in-cell simulations on exascale-class supercomputers},
\newblock in: \bibinfo{booktitle}{SC22: International Conference for High Performance Computing, Networking, Storage and Analysis}, \bibinfo{publisher}{IEEE}, \bibinfo{year}{2022}, pp. \bibinfo{pages}{25--36}.
%Type = Article
\bibitem[{Burau et~al.(2010)Burau, Widera, H{\"o}nig, Juckeland, Debus, Kluge, Schramm, Cowan, Sauerbrey, and Bussmann}]{burau2010picongpu}
\bibinfo{author}{H.~Burau}, \bibinfo{author}{R.~Widera}, \bibinfo{author}{W.~H{\"o}nig}, \bibinfo{author}{G.~Juckeland}, \bibinfo{author}{A.~Debus}, \bibinfo{author}{T.~Kluge}, \bibinfo{author}{U.~Schramm}, \bibinfo{author}{T.~E. Cowan}, \bibinfo{author}{R.~Sauerbrey}, \bibinfo{author}{M.~Bussmann},
\newblock \bibinfo{title}{{PIConGPU}: A fully relativistic particle-in-cell code for a {GPU} cluster},
\newblock \bibinfo{journal}{IEEE Transactions on Plasma Science} \bibinfo{volume}{38} (\bibinfo{year}{2010}) \bibinfo{pages}{2831--2839}.
%Type = Inproceedings
\bibitem[{Zenker et~al.(2016)Zenker, Widera, Huebl, Juckeland, Kn{\"u}pfer, Nagel, and Bussmann}]{zenker2016picongpu}
\bibinfo{author}{E.~Zenker}, \bibinfo{author}{R.~Widera}, \bibinfo{author}{A.~Huebl}, \bibinfo{author}{G.~Juckeland}, \bibinfo{author}{A.~Kn{\"u}pfer}, \bibinfo{author}{W.~E. Nagel}, \bibinfo{author}{M.~Bussmann},
\newblock \bibinfo{title}{Performance-portable many-core plasma simulations: Porting {PIConGPU} to {OpenPower} and beyond},
\newblock in: \bibinfo{booktitle}{High Performance Computing}, volume \bibinfo{volume}{9945} of \textit{\bibinfo{series}{Lecture Notes in Computer Science}}, \bibinfo{publisher}{Springer International Publishing}, \bibinfo{year}{2016}, pp. \bibinfo{pages}{293--301}.
%Type = Article
\bibitem[{Derouillat et~al.(2018)Derouillat, Beck, P{\'e}rez, Vinci, Chiaramello, Grassi, Fl{\'e}, Bouchard, Plotnikov, Aunai, Dargent, Riconda, and Grech}]{derouillat2018smilei}
\bibinfo{author}{J.~Derouillat}, \bibinfo{author}{A.~Beck}, \bibinfo{author}{F.~P{\'e}rez}, \bibinfo{author}{T.~Vinci}, \bibinfo{author}{M.~Chiaramello}, \bibinfo{author}{A.~Grassi}, \bibinfo{author}{M.~Fl{\'e}}, \bibinfo{author}{G.~Bouchard}, \bibinfo{author}{I.~Plotnikov}, \bibinfo{author}{N.~Aunai}, \bibinfo{author}{J.~Dargent}, \bibinfo{author}{C.~Riconda}, \bibinfo{author}{M.~Grech},
\newblock \bibinfo{title}{{Smilei}: A collaborative, open-source, multi-purpose particle-in-cell code for plasma simulation},
\newblock \bibinfo{journal}{Computer Physics Communications} \bibinfo{volume}{222} (\bibinfo{year}{2018}) \bibinfo{pages}{351--373}.
%Type = Article
\bibitem[{Diederichs et~al.(2022)Diederichs, Benedetti, Huebl, Lehe, Myers, Sinn, Vay, Zhang, and Th{\'e}venet}]{diederichs2022hipace}
\bibinfo{author}{S.~Diederichs}, \bibinfo{author}{C.~Benedetti}, \bibinfo{author}{A.~Huebl}, \bibinfo{author}{R.~Lehe}, \bibinfo{author}{A.~Myers}, \bibinfo{author}{A.~Sinn}, \bibinfo{author}{J.-L. Vay}, \bibinfo{author}{W.~Zhang}, \bibinfo{author}{M.~Th{\'e}venet},
\newblock \bibinfo{title}{{HiPACE++}: A portable, 3d quasi-static particle-in-cell code},
\newblock \bibinfo{journal}{Computer Physics Communications} \bibinfo{volume}{278} (\bibinfo{year}{2022}) \bibinfo{pages}{108421}.
%Type = Article
\bibitem[{Zhang et~al.(2021)Zhang, Myers, Gott, Almgren, and Bell}]{zhang2021amrex}
\bibinfo{author}{W.~Zhang}, \bibinfo{author}{A.~Myers}, \bibinfo{author}{K.~Gott}, \bibinfo{author}{A.~Almgren}, \bibinfo{author}{J.~Bell},
\newblock \bibinfo{title}{{AMReX}: Block-structured adaptive mesh refinement for multiphysics applications},
\newblock \bibinfo{journal}{The International Journal of High Performance Computing Applications} \bibinfo{volume}{35} (\bibinfo{year}{2021}) \bibinfo{pages}{508--526}.
%Type = Misc
\bibitem[{{FBPIC contributors}(2026)}]{fbpic2026docs}
\bibinfo{author}{{FBPIC contributors}}, \bibinfo{title}{{FBPIC} 0.27.0 documentation}, \bibinfo{year}{2026}. \URLprefix \url{https://fbpic.github.io/}, \bibinfo{note}{accessed 2026-09-03}.
%Type = Misc
\bibitem[{Sinn(2025)}]{sinn2025fbpicamd}
\bibinfo{author}{A.~Sinn}, \bibinfo{title}{{[DO NOT MERGE]} test on {AMD GPUs}}, \bibinfo{howpublished}{fbpic/fbpic Pull Request \#740}, \bibinfo{year}{2025}. \URLprefix \url{https://github.com/fbpic/fbpic/pull/740}, \bibinfo{note}{open, unmerged experimental pull request; head commit e3e299bc3d5de76c3bb8781b8eb70665206b149a; accessed 2026-09-03}.
%Type = Misc
\bibitem[{{FBPIC contributors}(2026)}]{fbpic2026parallel}
\bibinfo{author}{{FBPIC contributors}}, \bibinfo{title}{Parallelization of {FBPIC}}, \bibinfo{year}{2026}. \URLprefix \url{https://fbpic.github.io/overview/parallelisation.html}, \bibinfo{note}{{FBPIC} 0.27.0 documentation; accessed 2026-09-03}.
%Type = Article
\bibitem[{Jalas et~al.(2017)Jalas, Dornmair, Lehe, Vincenti, Vay, Kirchen, and Maier}]{jalas2017psatd}
\bibinfo{author}{S.~Jalas}, \bibinfo{author}{I.~Dornmair}, \bibinfo{author}{R.~Lehe}, \bibinfo{author}{H.~Vincenti}, \bibinfo{author}{J.-L. Vay}, \bibinfo{author}{M.~Kirchen}, \bibinfo{author}{A.~R. Maier},
\newblock \bibinfo{title}{Accurate modeling of plasma acceleration with arbitrary order pseudo-spectral particle-in-cell methods},
\newblock \bibinfo{journal}{Physics of Plasmas} \bibinfo{volume}{24} (\bibinfo{year}{2017}) \bibinfo{pages}{033115}.
%Type = Misc
\bibitem[{{NVIDIA Corporation}(2026)}]{nvidia2026cuda}
\bibinfo{author}{{NVIDIA Corporation}}, \bibinfo{title}{{CUDA C++ Programming Guide}}, \bibinfo{year}{2026}. \URLprefix \url{https://docs.nvidia.com/cuda/cuda-programming-guide/}, \bibinfo{note}{accessed 2026-09-03}.
%Type = Misc
\bibitem[{{Advanced Micro Devices, Inc.}(2026)}]{amd2026hipextensions}
\bibinfo{author}{{Advanced Micro Devices, Inc.}}, \bibinfo{title}{{HIP C++ Language Extensions}}, \bibinfo{year}{2026}. \URLprefix \url{https://rocm.docs.amd.com/projects/HIP/en/develop/how-to/hip_cpp_language_extensions.html}, \bibinfo{note}{accessed 2026-09-03}.
%Type = Article
\bibitem[{Ryoo et~al.(2008)Ryoo, Rodrigues, Stone, Stratton, Ueng, Baghsorkhi, and Hwu}]{ryoo2008carving}
\bibinfo{author}{S.~Ryoo}, \bibinfo{author}{C.~I. Rodrigues}, \bibinfo{author}{S.~S. Stone}, \bibinfo{author}{J.~A. Stratton}, \bibinfo{author}{S.-Z. Ueng}, \bibinfo{author}{S.~S. Baghsorkhi}, \bibinfo{author}{W.-m.~W. Hwu},
\newblock \bibinfo{title}{{Program optimization carving for GPU computing}},
\newblock \bibinfo{journal}{Journal of Parallel and Distributed Computing} \bibinfo{volume}{68} (\bibinfo{year}{2008}) \bibinfo{pages}{1389--1401}.
%Type = Misc
\bibitem[{{FBPIC contributors}(2026)}]{fbpic2026cuda}
\bibinfo{author}{{FBPIC contributors}}, \bibinfo{title}{{FBPIC GPU utility source: fbpic/utils/cuda.py}}, \bibinfo{howpublished}{GitHub source file at commit 2c4b9ca4c20847672fb7a86d61df3aeb766a9e8d}, \bibinfo{year}{2026}. \URLprefix \url{https://github.com/fbpic/fbpic/blob/2c4b9ca4c20847672fb7a86d61df3aeb766a9e8d/fbpic/utils/cuda.py}, \bibinfo{note}{accessed 2026-07-20}.
%Type = Article
\bibitem[{Hong and Kim(2009)}]{hong2009gpu}
\bibinfo{author}{S.~Hong}, \bibinfo{author}{H.~Kim},
\newblock \bibinfo{title}{{An analytical model for a GPU architecture with memory-level and thread-level parallelism awareness}},
\newblock \bibinfo{journal}{ACM SIGARCH Computer Architecture News} \bibinfo{volume}{37} (\bibinfo{year}{2009}) \bibinfo{pages}{152--163}.
%Type = Misc
\bibitem[{{Advanced Micro Devices, Inc.}(2026)}]{amd2026hipperformance}
\bibinfo{author}{{Advanced Micro Devices, Inc.}}, \bibinfo{title}{{HIP Performance Guidelines}}, \bibinfo{year}{2026}. \URLprefix \url{https://rocm.docs.amd.com/projects/HIP/en/latest/how-to/performance_guidelines.html}, \bibinfo{note}{accessed 2026-09-03}.
%Type = Misc
\bibitem[{Fanfarillo and Curtis(2023)}]{fanfarillo2023registerpressure}
\bibinfo{author}{A.~Fanfarillo}, \bibinfo{author}{N.~Curtis}, \bibinfo{title}{Register pressure in {AMD CDNA2} gpus}, \bibinfo{year}{2023}. \URLprefix \url{https://gpuopen.com/learn/amd-lab-notes/amd-lab-notes-register-pressure-readme/}, \bibinfo{note}{originally published 2023-05-17; updated 2024-06-26; accessed 2026-09-03}.
%Type = Incollection
\bibitem[{Hu et~al.(2021)Hu, Han, Han, and Shang}]{hu2021threadblock}
\bibinfo{author}{W.~Hu}, \bibinfo{author}{L.~Han}, \bibinfo{author}{P.~Han}, \bibinfo{author}{J.~Shang},
\newblock \bibinfo{title}{{Automatic Thread Block Size Selection Strategy in GPU Parallel Code Generation}},
\newblock in: \bibinfo{booktitle}{Parallel Architectures, Algorithms and Programming}, \bibinfo{publisher}{Springer Singapore}, \bibinfo{year}{2021}, pp. \bibinfo{pages}{390--404}.
%Type = Article
\bibitem[{Pereira et~al.(2020)Pereira, Pinheiro, and Schirru}]{pereira2020block}
\bibinfo{author}{C.~M. Pereira}, \bibinfo{author}{A.~L. Pinheiro}, \bibinfo{author}{R.~Schirru},
\newblock \bibinfo{title}{{Automatic block dimensioning on GPU-accelerated programs through particle swarm optimization}},
\newblock \bibinfo{journal}{Information and Software Technology} \bibinfo{volume}{123} (\bibinfo{year}{2020}) \bibinfo{pages}{106299}.
%Type = Incollection
\bibitem[{Lurati et~al.(2024)Lurati, Heldens, Sclocco, and van Werkhoven}]{lurati2024hip}
\bibinfo{author}{M.~Lurati}, \bibinfo{author}{S.~Heldens}, \bibinfo{author}{A.~Sclocco}, \bibinfo{author}{B.~van Werkhoven},
\newblock \bibinfo{title}{{Bringing Auto-Tuning to HIP: Analysis of Tuning Impact and Difficulty on AMD and Nvidia GPUs}},
\newblock in: \bibinfo{booktitle}{Euro-Par 2024: Parallel Processing}, \bibinfo{publisher}{Springer Nature Switzerland}, \bibinfo{year}{2024}, pp. \bibinfo{pages}{91--106}.
%Type = Article
\bibitem[{Stantchev et~al.(2008)Stantchev, Dorland, and Gumerov}]{stantchev2008particlegrid}
\bibinfo{author}{G.~Stantchev}, \bibinfo{author}{W.~Dorland}, \bibinfo{author}{N.~Gumerov},
\newblock \bibinfo{title}{{Fast parallel Particle-To-Grid interpolation for plasma PIC simulations on the GPU}},
\newblock \bibinfo{journal}{Journal of Parallel and Distributed Computing} \bibinfo{volume}{68} (\bibinfo{year}{2008}) \bibinfo{pages}{1339--1349}.
%Type = Article
\bibitem[{Kong et~al.(2011)Kong, Huang, Ren, and Decyk}]{kong2011gpupic}
\bibinfo{author}{X.~Kong}, \bibinfo{author}{M.~C. Huang}, \bibinfo{author}{C.~Ren}, \bibinfo{author}{V.~K. Decyk},
\newblock \bibinfo{title}{Particle-in-cell simulations with charge-conserving current deposition on graphic processing units},
\newblock \bibinfo{journal}{Journal of Computational Physics} \bibinfo{volume}{230} (\bibinfo{year}{2011}) \bibinfo{pages}{1676--1685}.
%Type = Misc
\bibitem[{{FBPIC contributors}(2026)}]{fbpic2026linearwakefield}
\bibinfo{author}{{FBPIC contributors}}, \bibinfo{title}{{FBPIC} linear-wakefield verification test: \texttt{tests/test\_linear\_wakefield.py}}, \bibinfo{year}{2026}. \URLprefix \url{https://github.com/fbpic/fbpic/blob/2c4b9ca4c20847672fb7a86d61df3aeb766a9e8d/tests/test_linear_wakefield.py}, \bibinfo{note}{gitHub source file at commit 2c4b9ca4c20847672fb7a86d61df3aeb766a9e8d; accessed 2026-09-04}.
%Type = Article
\bibitem[{Esarey et~al.(2009)Esarey, Schroeder, and Leemans}]{esarey2009laser}
\bibinfo{author}{E.~Esarey}, \bibinfo{author}{C.~B. Schroeder}, \bibinfo{author}{W.~P. Leemans},
\newblock \bibinfo{title}{Physics of laser-driven plasma-based electron accelerators},
\newblock \bibinfo{journal}{Reviews of Modern Physics} \bibinfo{volume}{81} (\bibinfo{year}{2009}) \bibinfo{pages}{1229--1285}.
%Type = Misc
\bibitem[{{FBPIC contributors}(2026)}]{fbpic2026lwfaexample}
\bibinfo{author}{{FBPIC contributors}}, \bibinfo{title}{{FBPIC} laser-wakefield acceleration example: \texttt{lwfa\_script.py}}, \bibinfo{year}{2026}. \URLprefix \url{https://github.com/fbpic/fbpic/blob/2c4b9ca4c20847672fb7a86d61df3aeb766a9e8d/tests/test_linear_wakefield.py}, \bibinfo{note}{gitHub source file at commit 2c4b9ca4c20847672fb7a86d61df3aeb766a9e8d; accessed 2026-09-04}.
%Type = Misc
\bibitem[{Vetter and Chambreau(2020)}]{vetter2020mpip}
\bibinfo{author}{J.~S. Vetter}, \bibinfo{author}{C.~Chambreau}, \bibinfo{title}{{mpiP}: A lightweight mpi profiler}, \bibinfo{year}{2020}. \URLprefix \url{https://software.llnl.gov/mpiP/}, \bibinfo{note}{mpiP 3.5 User Guide; accessed 2026-09-04}.

\end{thebibliography}



\end{document}